\documentclass[showpacs,12pt,amsmath,amssymb,pre]{revtex4}

\usepackage[T1]{fontenc}
\usepackage[cp1250]{inputenc}
\usepackage{graphicx}
\usepackage{placeins}
\usepackage{amsmath}
\usepackage{dsfont}

\newcommand{\re}{\mbox{Re}}

\newcommand{\bee}{\begin{eqnarray}}
\newcommand{\eee}{\end{eqnarray}}
\newcommand{\be}{\begin{equation}}
\newcommand{\ee}{\end{equation}}
\newcommand{\bifi}{\begin{figure}}
\newcommand{\efi}{\end{figure}}

\begin{document}
\title{Fingered growth in channel geometry: A Loewner equation approach.}
\author{T. Gubiec and P. Szymczak}
\affiliation{Faculty of Physics, Warsaw University, Ho\.za 69,
02-681 Warsaw, Poland}

\begin{abstract}
A simple model of Laplacian growth is considered, in which
the growth takes place only at the tips of long, thin fingers.
In a recent paper, Carleson and Makarov used the deterministic Loewner
equation to describe the evolution of such a system.
We extend their approach to a channel geometry and show that the
presence of the side walls has a significant influence on the evolution
of the fingers and the dynamics of the screening process,
in which longer fingers suppress the growth of the shorter ones.
\end{abstract}

\maketitle

\section{Introduction}

A variety of natural growth processes, including electrodeposition, viscous fingering,
solidification, dielectric breakdown or even growth of bacterial colonies can be modeled
in terms of Laplacian growth. In this model a plane interface between two phases moves
with velocity driven by a scalar field $u({\bf r},t)$ satisfying the Laplace equation
\bee
	\nabla^2 u({\bf r},t) = 0,
\label{lapl}
\eee
with the boundary condition $u({\bf r},t)=0$ at the phase interface. The normal velocity
of
the growing phase is proportional to the field gradient at the interface (or to some
power ($\eta$) of the gradient)
\bee
	v \sim \left|\nabla u({\bf r},t)\right|^{\eta}.
\label{veloc}
\eee
The harmonic field $u$ can represent e.g. temperature, pressure, or concentration,
depending
on the problem studied. An important property of Laplacian growth processes is the
{Mullins-Sekerka} instability of the advancing interface. The field gradient over a small
bump is  larger than that over the plane interface, thus, for $\eta > 0$, the bump grows
faster than adjacent areas of the interface and develops
into a finger. Contrastingly, for $\eta < 0$, the bumps are flattened and the growth is
stable.

Initial phases of the evolution of a plane interface are well understood in terms of
linear stability analysis, which yields the wavelength of the most unstable
perturbation~\cite{Vicsek,Meakin}. However, the later stages of the evolution are no
longer linear and hard to tackle analytically. Here we consider a simplified model of
the developed nonlinear state.
It is assumed that a number of {finger-like} protrusions were already formed (as a result
of
the initial
instability of the front) and the further growth of those fingers is assumed to take place
only at their tips (see Fig.~\ref{fig:iglowe}), with velocities proportional to the field
gradient. This dynamics is
deterministic: once the initial
geometry is given, the state of the system at any later time is uniquely determined.
Thus, such a model can be used when, except for the initial instability leading to the
finger
formation, the role of a noise in the evolution of the system can be neglected.
Additionally, we neglect another {noise-driven phenomenon}: the {tip-splitting} effect
when single finger bifurcates into two or more daughter branches.

\bifi
\begin{minipage}{0.5\linewidth}
\centering
\includegraphics[scale=1.8]{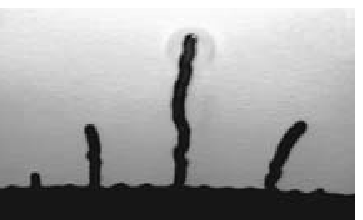}
\end{minipage}%
\begin{minipage}{0.5\linewidth}
\centering
\includegraphics[scale=0.5]{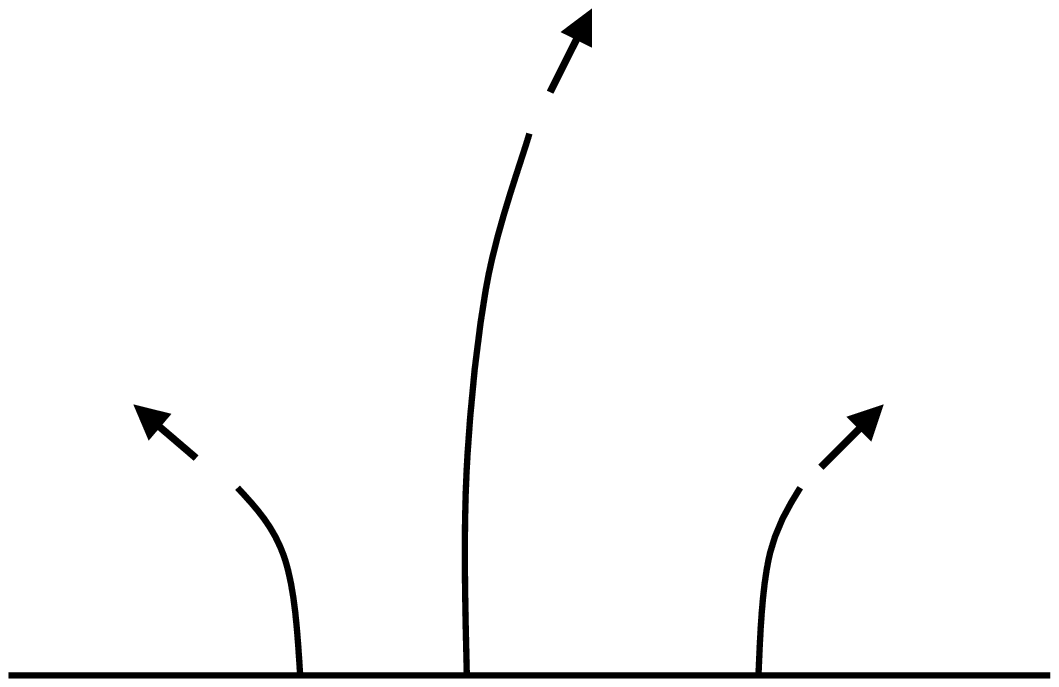}
\end{minipage}
\caption{Fingering in the combustion experiments~\cite{Zik:1999} and the
 theoretical model (right).}
	\label{fig:iglowe}

\efi

The above model of finger growth was formulated mathematically by Carleson and Makarov
\cite
{CarlesonMakarov:2002} (see also Selander's thesis \cite{Selander:1999}) and called by
them ,,the geodesic
Laplacian path model''. Independently, a similar idea was considered by Hastings in \cite{
Hastings:2001}.

In several experimental and numerical studies the patterns relevant to the
{above-introduced} model have been observed. The examples include dendritic growth in some
of the electrochemical deposition experiments~\cite{Kuhn:1994,Meakin},
channeling in dissolving rocks~\cite{Szymczak:2006} , {side-branches}
growth in crystallization \cite{Couder:1990,Couder:2005}, or fiber and microtubule growth~
\cite{Dogterom1993a,Dogterom1993b}.
Among the most beautiful experiments on the fingered growth are the combustion studies by
Zik and Moses~\cite{Zik:1998,Zik:1999}. In
those experiments (some of the results of which are reproduced in Figs.~\ref{fig:iglowe}
and~\ref{compet2}) a solid fuel is burnt in a
{Hele-Shaw} cell, i.e. in the narrow gap between two parallel plates. Near the flame
extinction, as the flux of the oxygen supplied to the system is being lowered, the initial
instability of the combustion front develops into the sparse fingers
(cf. Fig.~\ref{fig:iglowe}), which appear to evolve in a regular or {near-regular} way.

\bifi
\includegraphics[width=0.25\textwidth]{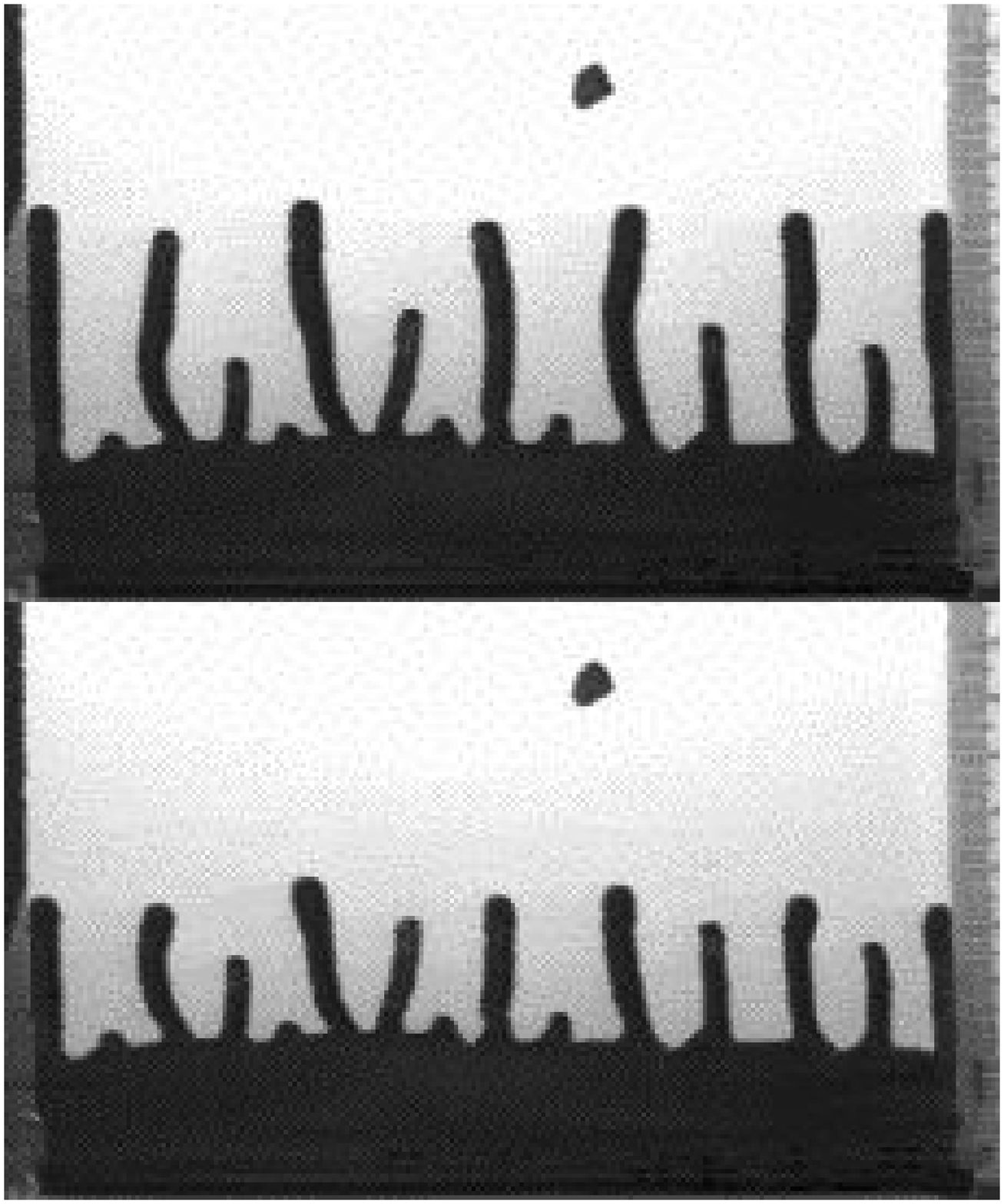}
	\caption{Competition in the fingered growth in combustion experiments~\cite{
Zik:1999}. Initially equal sized fingers (bottom) evolve towards
state where every other finger stops growing (top).
}
	\label{compet2}
\efi

\bifi
	\framebox{\includegraphics[width=0.35\textwidth]{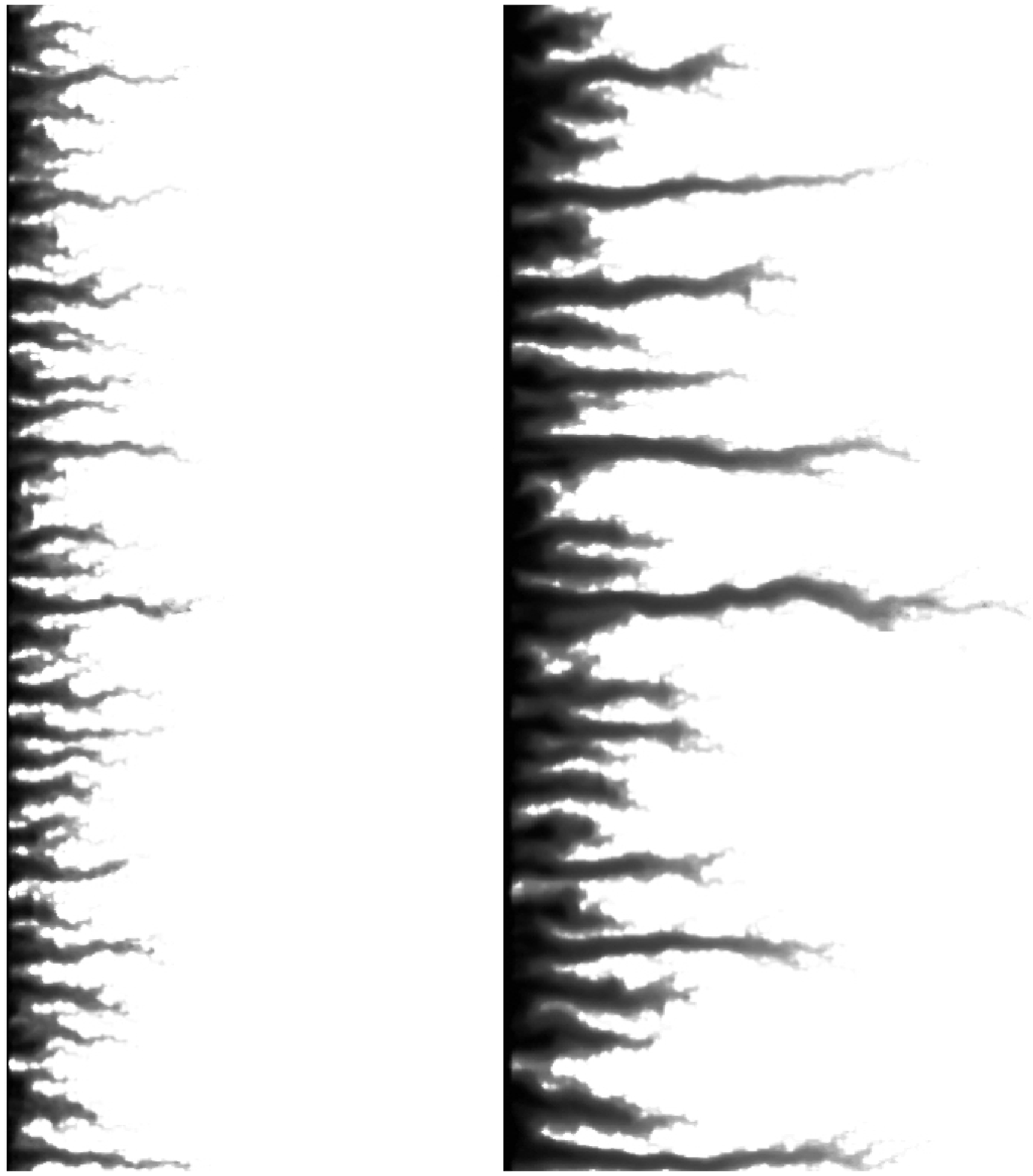}}
	\caption{Competitive dynamics of the channels in the dissolving rock fracture.
The figures present the dissolution patterns at two different time points.
Similarly to Fig.~\ref{compet2}, longer channels grow faster and suppress the growth of
the shorter ones
\cite{Szymczak:2006}.}
	\label{compet1}
\efi

Growth of the fingers demonstrates analogous instabilities as
the plane front, since the gradients around the tips of the longer fingers are larger than
those around shorter fingers. This leads to {so-called} ``shadowing effect'' - the longer
fingers grow faster and suppress the growth of the shorter ones in their neighborhood,
which in some cases gives rise to a {scale-invariant} distribution of finger lengths
in the {long-time} limit~\cite{Krug:1993,Huang:1997,Szymczak:2006}.
Figs.~\ref{compet2} and ~\ref{compet1} present examples of such a competitive growth in
the channeling processes in dissolving rock and in the combustion experiments described
above.

Most of the experiments on nonequilibrium growth mentioned above were performed in a
{quasi-2d} geometry. A convenient way of solving the Laplace equation in two
dimensions is to use a conformal mapping which transforms a domain under consideration to
some simpler region where the solution may easily be found.
A remarkable idea, due to Loewner~\cite{Loewner}, is to trace the evolution of the mapping
instead of the evolution of the boundary itself. It turns out, namely, that the evolution
of the map may be in many cases described by the first order ordinary differential
equation (Loewner
equation), which represents a considerable simplification in comparison to the partial
differential equation describing boundary evolution.
Loewner evolutions are intensely studied in the theory of univalent functions
(for general references see the monographs~\cite{Duren1983,Rosenblum:1994} and for a
recent
physical introduction see ~\cite{Gruzberg:2004,Bauer:2006}). The subject has recently
attracted a lot of attention in the
statistical physics community in context of Stochastic Loewner Evolution (SLE), which has
proved to be an important tool in the study of {two-dimensional} critical
systems~\cite{Lawler:2004,Kager:2004b,Lawler:2005,Bauer:2003}.

The exact form of Loewner equation depends on the shape of the domain in which the growth
takes place. Usually, it is either the complex {half-plane} (where the initial phase
boundary
corresponds to the real axis) or radial geometry (where the initial boundary is the
unit
circle in the complex plane).
 However, many experiments on the unstable growth are conducted in a channel geometry,
between two reflecting walls. In this paper we show how to extend the Carleson and Makarov
model to such a geometry. Using conformal mapping formalism we derive  Loewner
equation for that case, which allows us to find the dynamics of the fingers and analyze
the shadowing process. As it turns out the influence of the walls is often crucial
for the dynamics of the fingers.

\section{The Model}
\label{themodel}
With this introduction, let us formulate the model of finger growth to be considered.
The growth takes place at the tips of a finite number, $n$, of infinitely thin fingers
$\Gamma_{i} (t)$
(disjoint Jordan arcs)
\bee
	\Gamma_{i} (t) \subset \mathds{W}, \ \ \ \  i=1, \ldots n,
\eee
where $\mathds{W}$ is the domain in which the growth takes place and where the Laplace
equation needs to
be solved. The fingers extend from the boundary of  $\mathds{W}$ towards its interior in
such a way
that $\Gamma_{i}(t) \subset \Gamma_{i}(t')$ for $t'>t$. Additionally, on both the phase
interface (along boundary of $\mathds{W}$ ) and
along the fingers the condition
$u({\bf r},t)=0$ is imposed.

Since the finger is assumed to be infinitely thin, there is a singularity in
a field gradient at its tip. Namely, at a small distance $r$ from the tip of $i$th finger,
the gradient takes the form
\begin{eqnarray}
\nabla u({\bf r},t) =
\frac{C_i(t)}{{2\sqrt r }}\left( {\cos (\theta /2){\bf{e}}_r  + \sin (\theta /2){\bf{e}}_
\theta
} \right),
\end{eqnarray}
where the coefficients $C_i(t)$ depend on lengths and shapes of all the fingers.
In the above, the origin of coordinates is located at the tip of the finger and the polar
axis is directed along it. Following Derrida and Hakim \cite{DerridaHakim:1992}, we
introduce a
small circle of radius $r_0$ around the tip and define the finger growth rate as the
integral of field
gradient over
the circle
\bee
v_i(t) = \oint \hat{n} \cdot \nabla u({\bf r},t) \  ds \ = 2 \sqrt{r_0} \ C_i(t).
\label{eq:v_int}
\eee
Note that if the field $u({\bf r},t)$ describes the concentration, then the above integral
corresponds to
the total particle flux through the circle. The parameter $r_0$ should be of the order
of the finger width; its exact value does not influence the dynamics as long
as we assume it
to be the same for each finger. In such a case, the factor $2 \sqrt{r_0}$ may be absorbed
into the
definition of time, and we take  $v_i(t)$ equal to $C_i(t)$ (or to $C_i(t)^{\eta}$ in
$\eta$
growth).

To solve the Laplace equation we construct a time dependent map $g_t$:
\bee
 g_{t} : \mathds{W} \backslash  \left(\Gamma_{1}(t) \cup \ldots \cup \Gamma_{n} (t)
\right)
\rightarrow \mathds{W}
\eee
together with its inverse, $f_t$:
\bee
 f_{t} : \mathds{W}   \rightarrow \mathds{W} \backslash  \left(\Gamma_{1}(t) \cup \ldots
\cup \Gamma_{n} (t) \right), \\
 f_{t} \circ g_{t} = g_{t} \circ f_{t} =id \ , \  t \geq 0,
\eee
as illustrated in Fig.~\ref{mapa}.
The function $f_t$ can be extended to a continuous function of the boundary,
which is {two-to-one} along the fingers, except of the tips where it is {one-to-one}.
Thus the tips of the fingers (denoted by $\gamma_i(t)$) may be added to the domain of the
function $g_t$ with the corresponding images, $a_i(t)$:
\bee
a_i=g_t(\gamma_i).
\eee

\bifi
	\includegraphics[width=0.75\textwidth]{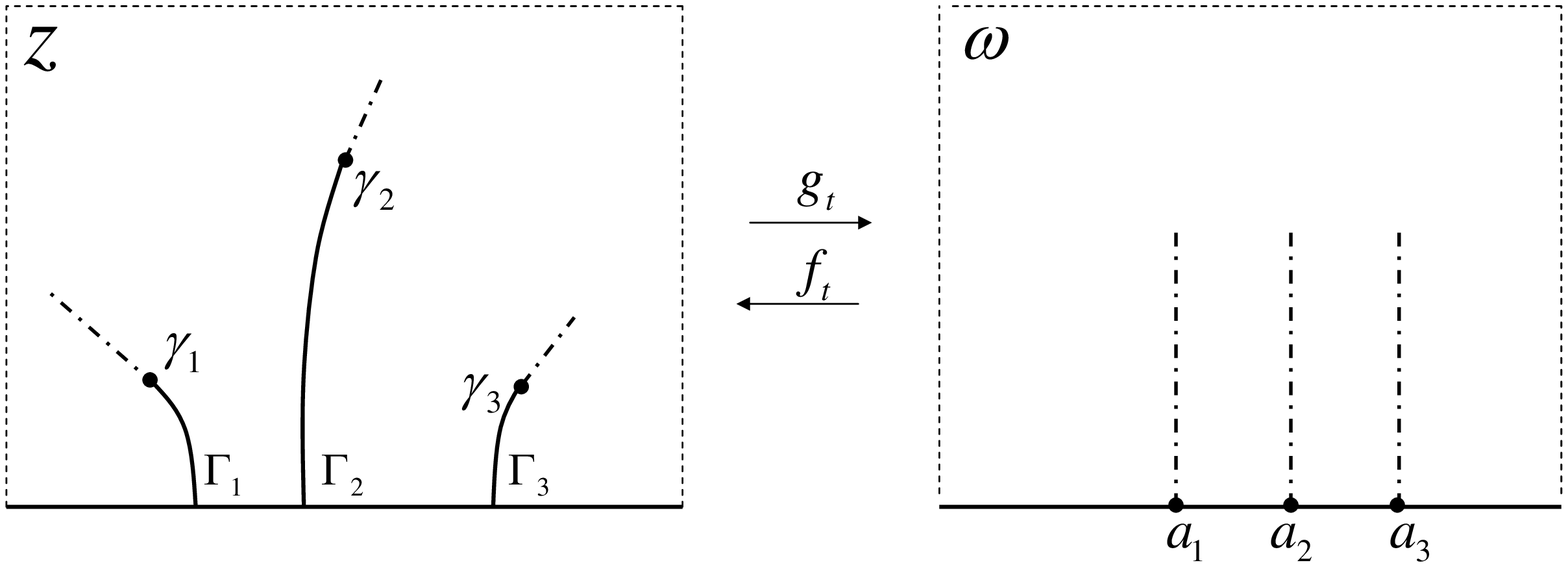}
	\caption{An example configuration of three fingers
($\Gamma_1, \Gamma_2, \Gamma_3$)
extending from the real axis in the physical plane ($z$-plane). The mapping $g_t$ maps the
exterior of the fingers onto the empty half-plane ($\omega$-plane). The images of the tips
$\gamma_i$ are located on the real line at the points $x=a_i$. The gradient lines of the
Laplacian field ({dot-dashed}) in the $z$-plane are mapped onto the vertical lines in the
$\omega$
plane. In a given moment of time, the fingers grow along the gradient lines the images of
which pass through the points $a_i$.}
	\label{mapa}
\efi

It may be shown (see Appendix~\ref{efbis} and also Refs.~\cite{
Selander:1999,CarlesonMakarov:2002,DerridaHakim:1992}) that the
growth rate of a finger (\ref{eq:v_int}) can be
expressed in terms of the map $f_t$ as
\bee
v_i(t) \sim \left| f''_t (a_i(t)) \right|^{-\eta/2}.
\label{eq:v_f}
\eee

As mentioned in the Introduction, the above model of finger growth was proposed in
Refs.~\cite{Selander:1999, CarlesonMakarov:2002} and independently
in~\cite{Hastings:2001}.
This model may be also looked upon as a deterministic generalization of Meakin and Rossi
\cite{rossi86,meakin86} needle growth model -- a simplified, {non-branching} version
of Witten and Sander's {diffusion-limited} aggregation (DLA) \cite{DLA}. In Meakin and
Rossi
model the growth is allowed
to occur only in the direction away from the substrate, which results
in a forest of parallel needles with a broad distribution of heights.
Cates \cite{Cates:86} obtained average density of the aggregate in such a system in mean
field approximation whereas Evertsz \cite{Evertsz:90} calculated fractal dimensions of
needle structures for different values of the exponent $\eta$.

Yet another modification of DLA was studied in {so-called} ``{polymer-growth model}''
\cite{Bradley:86,Debierre:86,Meakin:88} where, as before, the particles were allowed to
attach to the tip of the growing chain only, but the condition that the growth must occur
in the direction away from the substrate was relaxed.
This time, the process resulted in a set of {fiber-like} chains of different lengths and
shapes.

The above described, {DLA-based} models are stochastic in nature. On the other hand, the
deterministic versions of the Meakin and Rossi needle growth model
were studied by several authors
\cite{
Peterson:1989,PetersonFerry:1989,Kurtze:1991,DerridaHakim:1992,Peterson:1998,Krug:1993,
Huang:1997,Bernard:2001,Sakaguchi:2007}, mostly by conformal mapping techniques, which
were first applied to {DLA-like} aggregates by Shraiman and Bensimon~\cite{Shraiman:1984},
 Ball~\cite{Ball:1986}, and Szep and Lugosi~\cite{SzepLugosi:1986}.
 The majority of authors considered the problem in a radial
geometry, with a set of straight needles growing  radially from the origin, whereas in
\cite{Krug:1993,Huang:1997,Bernard:2001} a set of parallel needles with
alternating lengths was analyzed in periodic boundary conditions. An important difference
between the above quoted needle models and the model considered here is that we do not
constrain the fingers to follow the straight line, instead they can bend in the direction
of the field gradient at the tip. Thus, whereas the needle models are {well-suited} to
describe a strongly anisotropic growth of rigid structures such as the sidebranching
dendrites in solidification, they are less suited to describe the phenomena in which the
growing structures may become deflected by the field, as it is the case in the
{above-described} combustion experiments, channel formation in the dissolving rock or in a
number
of {fibril-growth} processes.

In the next section, we sketch the derivation of the Loewner equation for the fingered
growth in the {half-plane}. In the context of the {Carleson-Makarov} model, this problem
was analyzed by Selander in
\cite{Selander:1999}. We repeat this derivation here, because it is the simplest case
which
illustrates the method involved, which will be then used to tackle the channel geometry
case.

\section{Fingered growth in the {half-plane}}\label{hplane}

In this section, following~\cite{Selander:1999}, we present the description of the
fingered growth in the complex {half-plane} in terms of the conformal mappings. The
solution
of the Laplace equation in the
empty {half-plane}, satisfying the boundary conditions: $u({\bf r})=0$ on the real axis
and $\frac{\partial u}{\partial y} \underset{y \rightarrow \infty}{\longrightarrow}  1$ (
constant flux at infinity), is given simply by
\bee
u(x,y)=y.
\label{hplanesol}
\eee
The map $g_t$ from the exterior of the fingers to the empty half-plane will be uniquely
determined by the {so-called} hydrodynamic normalization at
infinity
\bee
	\lim_{z \rightarrow \infty} g_{t}(z)-z = 0,
	\label{eq:normalizacja}
\eee
which ensures that the flux at infinity is unaffected by boundary movements.
One of the possible ways of finding the map $g_t$ is to construct it as a
composition of slit mappings, with the slit length going to 0. In the case of the
upper {half-plane} $\mathds{C}_{+}= \left\{z \in \mathds{C}: \re (z)>0\right\}$.
the slit mapping
\bee
	\phi : \mathds{C}_{+} \backslash  \left( a+i \left[ 0,h \right] \right)
\rightarrow \mathds{C}_{+}, \ \ \ \ \ h>0
\eee
is of the form
\bee
	\phi(z)=\sqrt{(z-a)^{2}+h^2}+a.
	\label{eq:slit}
\eee
In Eq.~\eqref{eq:slit} the square root is specified by demanding that
$\phi(z) \rightarrow z$
at
infinity, thus the difference $(\phi(x)-a)$ is negative on the real axis for $x<a$ and
positive for
$x>a$.
As
stated above, there is a branch cut along $\left( a+i \left[ 0,h \right] \right)$; the
left
and right side of this segment are mapped on the interval $[-h, 0]$ and $[0,h]$
respectively.

In the simplest case of the single finger ($n=1$) the idea of the construction of the
mapping by the composition of successive slit maps is shown in Figure
\ref{composition}.
\bifi[ht]
	\centering
		\includegraphics[width=0.4\textwidth]{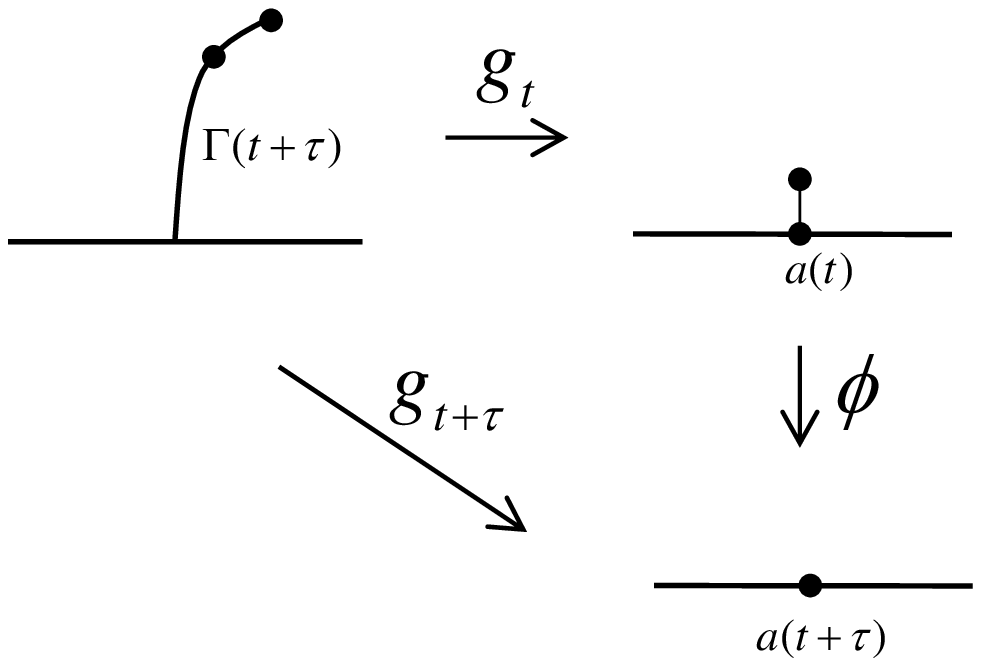}
	\caption{Illustration of the composition of conformal maps described in the text.
The mapping $g_{t+\tau}$ is obtained as the composition of $g_t$ and the elementary slit
mapping~\eqref{eq:slit}.}
	\label{composition}
\efi
The map $g_{t}$ maps the arc $\Gamma(t)$ into the real axis $\mathds{R}$, while
$g_{t+\tau}$ maps the longer arc $\Gamma(t+\tau)$ into $\mathds{R}$. The image of the tip,
$a$, is also in general time-dependent, $a=a(t)$. For small $\tau$,
the mapping
$g_{t+\tau}$ may be obtained by the composition
\bee
 g_{t+\tau}=\phi \circ g_{t} + O(\tau^{2}),
 \label{eq:zlozenie}
\eee
where $\phi (z)$ is the above defined slit mapping~\eqref{eq:slit}. Since the growth
rate of
the finger is not constant, the slit length, $h$, will be a function of time. It is
convenient to choose $h$ in the
form
\begin{equation}
h=\sqrt{2 \tau d(t)},
\label{hd}
\end{equation}
where $d(t)$ is the {so-called} growth factor~\cite{Selander:1999}. The square root
dependence
of the slit length on the
timestep ensures that the increment of length of the finger will be linear in $\tau$. The
slit mapping $\phi (z)$
is thus given by
\bee
 \phi(z)=\sqrt{(z-a(t))^{2}+2 \tau d(t)}+ a(t).
\label{phiz}
\eee
In the composition (\ref{eq:zlozenie}) only the terms linear in $\tau$ are needed,
hence
\bee
 g_{t+\tau}=\phi \circ g_{t}= g_{t} + \frac{ \tau d(t)}{g_{t}-a(t)}  + O(\tau^
{
2}), \\
 \frac{g_{t+\tau}-g_{t}}{\tau}= \frac{ d(t)}{g_{t}-a(t)} +O(\tau).
\eee
In the limit $\tau \rightarrow 0$ we obtain the Loewner equation for one finger in the
{half-plane}
\bee
	\dot{g}_{t}= \frac{ d(t)}{g_{t}-a(t)},
	\label{eq:loewner}
\eee
with the initial condition
\bee
g_0(z) = z
\eee
corresponding to the empty space with no fingers. Note that the pole of the equation~
\eqref{eq:loewner} is located at
the image of the tip, $a(t)=g_t(\gamma)$.
To relate the growth factor $d(t)$ to the finger velocity let us notice that for a
finite
$\tau$, the composition $\phi \circ g_{t}$ increases the length of the finger
approximately by
$\tau d(t) \left| f''_{t}(a(t)) \right| $, thus the growth velocity is given by
\bee
 v(t) = d(t)  \left| f''_{t}(a(t)) \right| .
\eee
Hence, using (\ref{eq:v_f}) we get
\bee
 d(t) = \left| f''_{t}(a(t)) \right|^{-\eta/2 -1}.
 \label{eq:dlaplace}
\eee
The final element of our description is the position of the pole $a$ as a function of
time.
This function may be found from the condition that the finger grows along the direction of
the gradient near the
tip. Due to the singularity at the tip, it is more convenient to work in
the $\omega$-plane.
Namely, the gradient lines in the physical plane ($z$-plane) are mapped by $g_t$ onto the
vertical lines
in the $\omega$-plane, as
illustrated in Fig.~\ref{mapa}. The counterimage of the veritcal line
$\omega=a+i s, \ s>0$ will thus define the growth
direction in the $z$ plane. In the case of a single finger, from symmetry, the gradient
line in the $z$ plane is also a vertical line -- the one passing through the tip of the
finger. Thus
in
this
case the
position of the
pole must be constant
\bee
a(t)=const.=a(0).
\eee

The Loewner equation can be generalized to the {$n$-finger} case by analyzing the
composition
of
$n$ slit mappings, $\phi_i$, one for each finger. Analogously to~\eqref{phiz}, the slit
mappings are then given by
\bee
	\phi_i(z)=\sqrt{(z-a_i(t))^{2}+2\tau d_i(t)}+a_i(t),
\eee
where $a_i(t)$ and $d_i(t)$ are the image of the tip and the growth factor of the ith
finger,
respectively.
 This leads to the Loewner equation of the form
\cite{Selander:1999,CarlesonMakarov:2002}
\bee
 \dot{g}_{t}= \sum^{n}_{i=1}\frac{ d_{i}(t)}{g_{t}-a_i (t)}.
 \label{eq:loewner1}
\eee
This time, however, to force the fingers to
grow along gradient lines, the functions $a_{i}(t)$ need to change in time, since the tip
images
$a_i=g(\gamma_i)$ are moved by slit mappings $\phi_j$ with $j \neq i$. This leads to the
following
condition
for the motion of the poles
\bee
 \dot{a}_{i}(t)=\sum^{n}_{ \stackrel{j=1}{j \neq i} } \frac{d_{j}(t)}{a_{i}(t)-a_{j}(t)}.
 \label{eq:a1}
\eee

Let us note at that point that the above idea of generating the growing aggregate by
iterated conformal maps was applied also to the original DLA problem in a seminal paper by
Hastings and Levitov~\cite{Hastings:1998}. In fact, the {``strike''-mapping} proposed by
them in~\cite{Hastings:1998}
is a counterpart of the slit mapping~\eqref{eq:slit} in the radial geometry. Analogous
constructions for the DLA in a channel and cylindrical geometry were proposed in
{Refs.~\cite{Somfai:2003,Taloni:2006}}. However, unlike the model considered here, those
models were
stochastic in nature, and - as usual in DLA - generated {noise-driven}, branched, fractal
structures.
A deterministic version of Hastings and Levitov construction was proposed by Hastings
in~\cite{Hastings:2001}. Although he did not use the formalism of the Loewner equation,
his model is essentially analogous to that presented above.

Going back to the system of equations (\ref{eq:loewner1}-\ref{eq:a1}),
let us now look at the solutions in a few simplest cases.
As mentioned, the single finger case is rather straightforward: the finger grows
vertically along the line $x=a$. A more interesting case is that of two fingers.
Here the results depend in a significant way on the value of the exponent $\eta$ used.
A related problem in a slightly different geometry
{(growth in $\mathds{W}\equiv \mathds{C} \setminus {\mathds{R}_+}$)} was considered by
Carleson and
Makarov in~\cite{CarlesonMakarov:2002}. Although their results are not directly applicable
here,  it
is
relatively straightforward to repeat their derivations for the growth in the half-plane.
Namely, it turns out that there are three regimes in the behavior of the fingers,
depending on
the value of the exponent $\eta$. For $-\infty< \eta<\eta_c \approx 0.43$, the two
fingers,
irrespectively of
their initial positions, grow symmetrically, as illustrated in Fig.~\ref{halfplane}. Next,
for $\eta_c < \eta < \eta_c^{\prime}=2/3$, the symmetric solution becomes unstable; one of
the fingers
starts to grow faster and screens the other. However, the screening is only partial and
the
ratio of finger velocities $v_1/v_2$ goes to a positive constant (different from 1) when
$t \rightarrow \infty$. Finally, for $\eta > \eta_c^{\prime}$ there is a stronger
screening and the
ratio of the velocity of the slower finger to that of the faster one goes to zero
asymptotically.

\bifi[ht]
	\centering
		\includegraphics[width=0.7\textwidth]{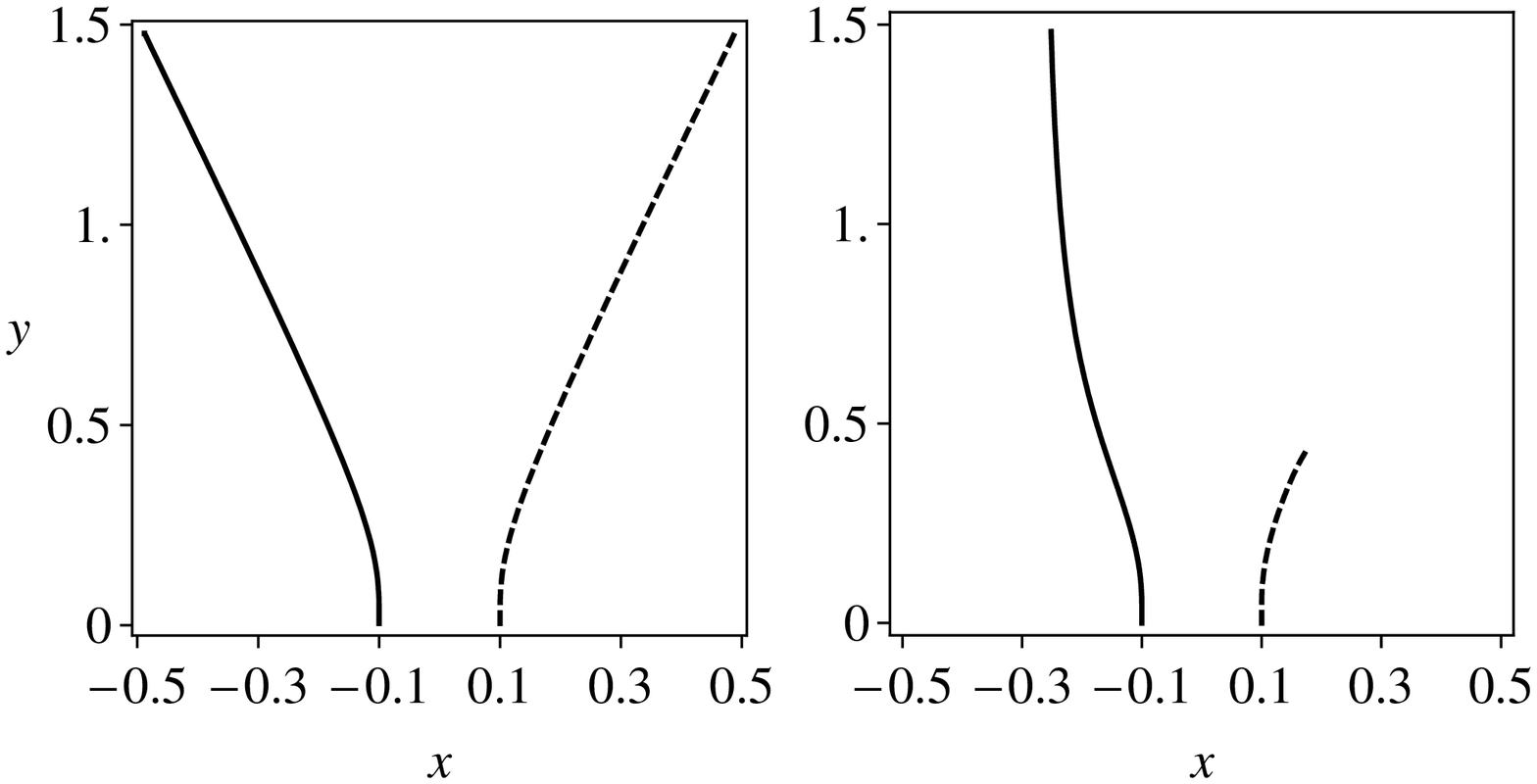}
		\caption{Two fingers growing in the half-plane for $\eta=-2$ (left panel)
and $\eta=4$ (right panel). The initial
positions of the fingers are $a_1(0)=-a_2(0)=0.1$.}
		\label{halfplane}
\efi

The shape of the fingers in the first regime may be obtained analytically. To this end, it
is most convenient to consider the case of $\eta=-2$ which, according to
Eq.~\eqref{eq:dlaplace}
corresponds to the evolution with constant growth factors, $d_i=d$. For further
analysis, we choose the coordinates in such a way that
$a_1(0)=-a_2(0)=a_0$.
Eq.~\eqref{eq:a1} may
be then readily integrated to yield
\bee
a_{1,2}(t)=\pm \sqrt{a_0^2+d_0t}.
\eee
Inserting the above into the Loewner equation~\eqref{eq:loewner1}, we can find $g_t$ and
then obtain
the positions of the tips $\gamma_i(t)$ as a function of time. They are given implicitly
by
\bee
\pm 16 (a_0^2 + d_0t)^{5/2} = \gamma_{1,2}(t) (\gamma_{1,2}(t)^2-5a_0^2)^2,
\label{traj}
\eee
where we take only the roots which lie in the upper {half-plane} and satisfty the initial
condition $\gamma_i(0)=a_i(0)$. An example trajectory with
$a_0=0.1$ is shown in Fig.~\ref{halfplane}. We see that the fingers repel each other,
which is due to the term $1/(a_j-a_i)$ in the evolution equation of
the poles~\eqref{eq:a1}. Asymptotically, as $t \rightarrow \infty$,
the relation~\eqref{traj} takes the form
\bee
\pm 16 (d_0t)^{5/2} = (\gamma_i(t))^{5},
\eee
from which it may be concluded that the fingers tend to the straight lines
$\text{arg}(z)=\frac{2 \pi}{5}$ and
$\text{arg}(z)=\frac{3 \pi}{5}$ with
the angle $\pi/5$ between them. Although the above result was obtained for $\eta=-2$ case,
as mentioned above, such a symmetric solution remains stable up to $\eta_c$. Thus, in this
range of $\eta$ the fingers follow the same trajectories as in Fig.~\ref{halfplane}, but
with
different velocities than those in $d=const.$ case. In fact, whenever the fingers grow
symmetrically, with equal growth factors, $d_1(t)=d_2(t)$, one may always rescale the time
coordinate by  defining $t' = \int^{t}_{0} d_1(t'') d t'' $ and thus reduce the problem to
the constant growth factor ($d_1=d_2=1$) case.

The right panel of Fig.~\ref{halfplane} shows the shape of the finger in the strongly
unstable case, $\eta=4$. In this case the solution was obtained numerically, as described
in Appendix~\ref{numer}. As it is observed, initially the tips follow the same trajectory
as in the stable case, but then due to the numerical noise in the computations, the
instability sets in, one of the
fingers outgrows the other and then continues along the vertical direction. The other
finger slows down and the ratio of its velocity to that of the winning finger goes to
zero.

The fact that there is a range of positive values of $\eta$ for which the symmetric
solution is stable seems surprising at first sight. However, in the unbounded domain the
fingers have the possibility to escape from each other to the regions where the influence
of another finger is smaller.
Such a behavior of the fingers may be hardly observed in a bounded system, because of the
wall effect which makes it impossible for the fingers to escape from each other. In the
next section, we show how to take into account the
presence of the side walls in the system and how their presence affects the dynamics of
the fingers.

\section{Fingered growth in the channel geometry}
In this section we consider the growth of the fingers in the channel geometry, i.e.
in the domain
\bee
	\mathds{P}=\left\{ z=x+iy\in \mathds{C}: y > 0 , \quad x \in \ ]-1,1[ \
\right\},
\eee
with the boundary conditions $u=0$ at a bottom wall $[-1,1]$
and along the fingers, and $\frac{\partial u}{\partial x} = 0$ on the sides (which
corresponds
to
reflecting boundary conditions at the impenetrable side walls) and
$\frac{\partial u}{\partial y} = 1$
at infinity (see Fig.~\ref{fig:pasek}).
\bifi
	\centering
	\includegraphics[width=0.4\textwidth]{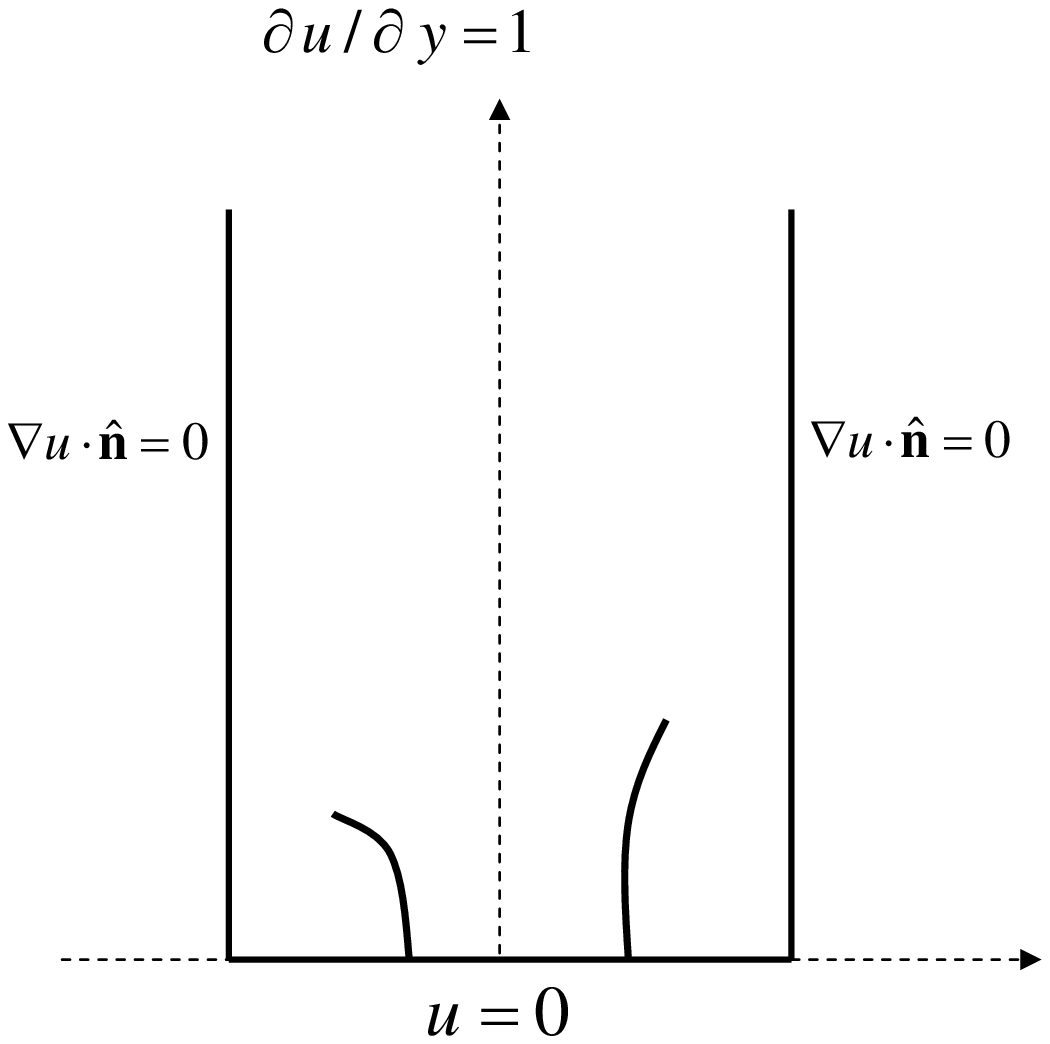}
	\caption{The geometry of the channel together with the boundary conditions}
	\label{fig:pasek}
\efi
For an empty channel with the above
boundary conditions, the solution of Laplace equation is again given by
\bee
	u(x,y)=y .
\eee
in full analogy to the half-plane case~\eqref{hplanesol}.

Our goal is to construct the mapping $g_t$
\bee
 g_{t} : \mathds{P} \backslash  \left(\Gamma_{1}(t) \cup \ldots \cup \Gamma_{n} (t)
\right)
\rightarrow \mathds{P}
\eee
together with its inverse $f_t$. The boundary condition at the bottom wall is different
from that at the sides, thus the
points
$-1$ and $1$ must remain fixed under the mapping $f_t$, or, in terms of $g_t$
\bee
	\lim_{z \rightarrow -1} g_t(z)+1 = \lim_{z \rightarrow 1} g_t(z) - 1 = 0 \label{
eq:normalizacjaP} .
\eee
Additionally, we require that $g_t$ keeps the point at infinity fixed, i.e.
\bee
\text{Im} \left( g_t(z)\right) \rightarrow \infty \ \ \ \ \ \text{for} \quad \text{Im}(z)
\rightarrow
\infty
\eee
 Those conditions define the conformal map $g_t$ uniquely.

First let us consider the case of the single finger. The derivation of Loewner equation in
this
case will be analogous to that presented above for the {half-plane}. However, the slit
mapping
is now different. The idea of construction of the mapping is shown in
Fig.~\ref{fig:zlozeniepasek}. As before, we begin with a short slit of length
$h=\sqrt{2 \tau d}$.
\bifi[ht]
	\centering
	\includegraphics[width=0.65\textwidth]{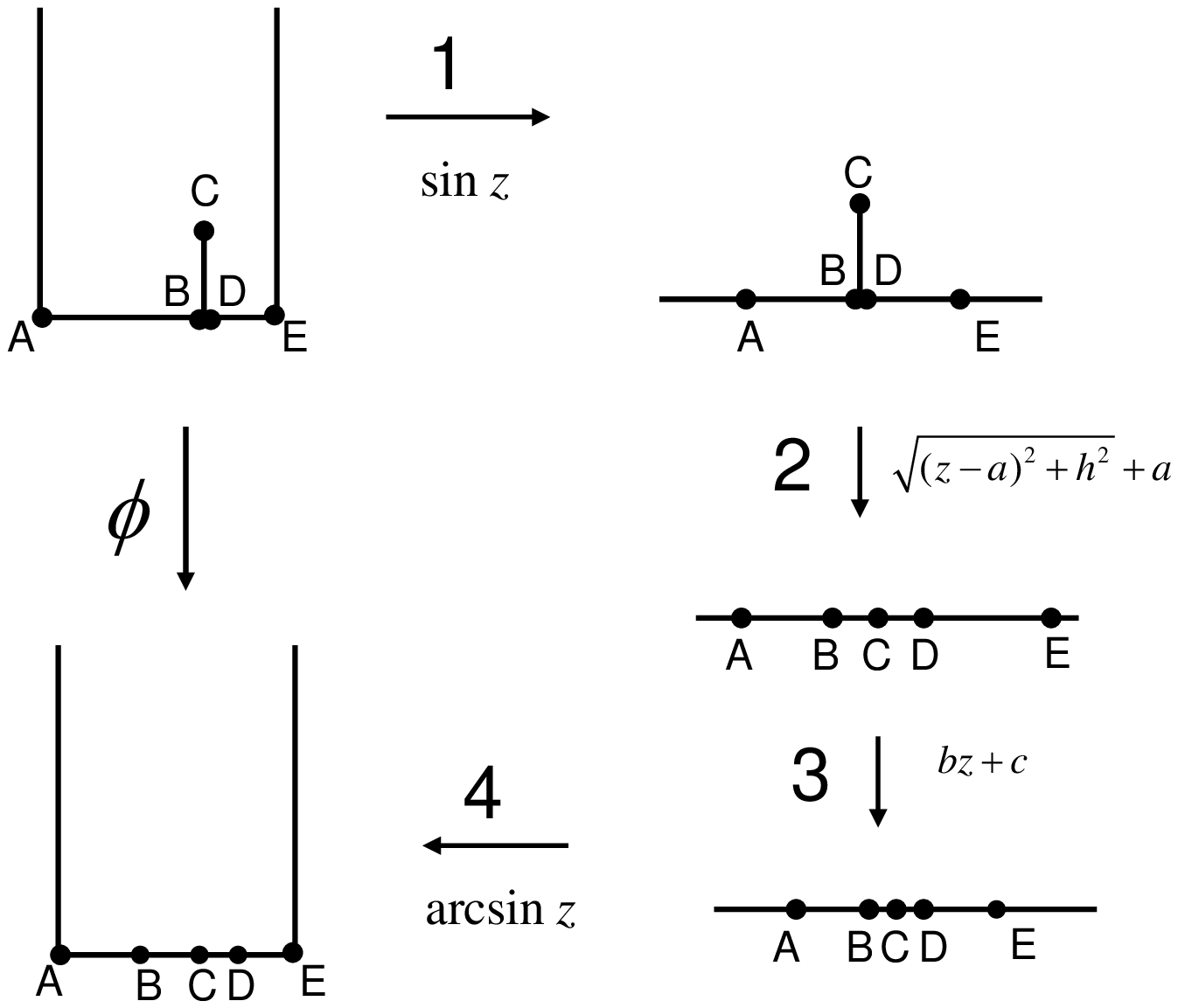}
	\caption{Schematic picture of the mappings used in construction of elementary slit
mapping  $\phi$ in the channel geometry}
	\label{fig:zlozeniepasek}
\efi
The first function `1' transforms
$\mathds{P}$ into the {half-plane} keeping points $A=-1$, $E=1$ fixed and is given by
$\sin \left(\frac{\pi}{2} z \right)$.  When the slit length is small,
$h << 1$, its length after the transformation `1' can be calculated from
the Taylor expansion up to linear terms in $\tau$
\begin{equation}
	\sin \left(\frac{\pi}{2} (a+i\sqrt{2 \tau d}) \right) \approx  \left( 1 + \frac
{\pi
^2}{4} \tau d \right)  \sin \left(\frac{\pi}{2} a \right) + \frac{i \pi}{2} \sqrt{2 \tau
d}
\  \cos \left(\frac{\pi}{2} a \right)+\ldots \quad .
\end{equation}
Denoting the new length of the slit by $\sqrt{2 \tau d'}$ we get
\bee
	 d'&=&d \frac{\pi^{2}}{4} \cos^{2}\left(\frac{\pi}{2}a\right) + O(\tau^2).
\eee
Function `2' is the slit mapping in the {half-plane} given by (\ref{eq:slit}).
The composition of
`1' and `2' reads then
\bee
	 \varphi (z) &=& \sqrt{\left( \sin\left(\frac{\pi}{2}z\right)-\beta \sin\left(
\frac
{\pi}{2}a\right)\right)^{2}+2 \tau d'},
\label{varphi}
\eee
where $\beta=\left( 1 + \frac{\pi^2}{4} \tau d \right)$.
The above slit mapping moves the points A and E. To shift them back to -1 and 1, we use
additional map, `3', which is a linear function with real parameters. The map `4' is
just the inverse of `1' , i.e. $ \frac{2}{\pi}\arcsin(z)$. The final form of $\phi$ is
thus given by
\bee
	\phi (z) =  \frac{2}{\pi} \arcsin\left[\frac{
2 \varphi(z) - (\varphi (1)+ \varphi (-1))}{ \varphi (1)-\varphi (-1)} \right].
	\label{eq:phiP}
\eee
Keeping only the terms linear in $\tau$
\bee
 \phi (z) = z + \tau d\frac{\pi}{2} \frac{\cos\left(\frac{\pi}{2}z\right)}{\sin\left(\frac
{
\pi}{2}z\right)-\sin\left(\frac{\pi}{2}a\right)} + O(\tau^{2}), \label{eq:rozwiniecieP}
\eee
we obtain Loewner equation of the form
\bee
		 \dot{g}_{t}  = d(t) \frac{\pi}{2} \frac{\cos\left(\frac{\pi}{2}
g_{t} \right)}{\sin\left(\frac{\pi}{2} g_{t} \right) - \sin\left(\frac{\pi}{2}a(t)
\right)}. \label{eq:loewnerP}
\eee
Note that due to the presence of the side walls the slit mapping is no longer symmetric,
in the
sense that the images of the points $B$ and $D$ at the base of the finger are asymmetric
with respect
to the image of C (cf. Fig.~\ref{fig:zlozeniepasek}). This means that, in contrast
to the
{half-plane} case, the pole $a=g(\gamma)$ will be shifted by the mapping. This shift may
be obtained
from the Loewner equation in the limit $g(z) \rightarrow a$. However, the equation is
singular at that point,
thus we take the symmetric limit from both sides towards a singularity:
\bee
	\dot{a}(t)= \lim_{\epsilon \rightarrow 0}  \frac{W(a - \epsilon)+W(a + \epsilon)
}{2},
\label{eq:azlozenia}
\eee
where $W(g)=d \frac{\pi}{2} \cos (\frac{\pi}{2}
g ) /\left( \sin (\frac{\pi}{2} g ) - \sin (\frac{\pi}{2}a
) \right)$.
Explicit evaluation of the limit yields
\bee
	\dot{a}(t)=- \frac{\pi}{4} d(t) \ \tan \left(\frac{\pi}{2}a(t)\right).
\label{eq:aprimP}
\eee
The equation for the {$n-$finger} case may be obtained, as before, by the composition of
$n$
slit mappings, one for each
finger, which leads to
\bee
		 \dot{g}_{t} =\frac{\pi}{2}  \sum_{i=1}^{n} d_{i}
\frac{\cos\left(\frac{\pi}{2} g_{t} \right)}{\sin\left(\frac{\pi}{2} g_{t} \right) - \sin
\left(\frac{\pi}{2}a_{i}\right)}.
\label{Loewnern}
\eee

Finally, to derive the condition for the motion of the poles in the {$n-$finger} case, we
need to add self terms of the form~\eqref{eq:aprimP} and the interaction terms, which may
be obtained from~\eqref{Loewnern} by taking
$g_t=a_j$ with $i \neq j$. This leads to
\begin{equation}
		 \dot{a}_{j} = - \frac{\pi}{4} d_{j}
\tan\left(\frac{\pi}{2}a_{j}\right)
 + \frac{\pi}{2}\sum^{n}_{ \stackrel{i=1}{i \neq j} } d_{i}  \frac{\cos\left(\frac{\pi}{
2} a_{j} \right)}{\sin
\left(\frac{\pi}{2} a_{j} \right) - \sin\left(\frac{\pi}{2}a_{i}\right)}.
\label{eq:aprimP_n}
\end{equation}

The presence of the self term,~\eqref{eq:aprimP}, attracts the pole to $a=0$, which is a
stable fixed point of Eq.~\eqref{eq:aprimP} and causes the finger to bend in the direction
of the centerline of the channel. It is most clearly seen in the single finger case. For
the constant growth factor, $d(t)=d_0$ (which corresponds to $\eta=-2$)
equation~\eqref{eq:aprimP} may be solved explicitly to yield
\bee
\sin \left( \frac{\pi}{2} a_{t} \right)= e^{-\frac{\pi^{2}d_0}{8} t} \sin\left(\frac{\pi}{
2}
a_{0}\right).
\eee
Thus,  $\sin (\frac{\pi}{2} a_{t})$ goes to zero exponentially. A corresponding trajectory
of
the finger may be
expressed implicitly through elliptic integrals and is shown
in Fig.~\ref{onefinger}. As it is observed, the finger starts at $z=a(0)$, initially grows
perpendicularly to the bottom wall, but very soon the influence of the walls becomes
important and the finger is attracted to the symmetric position in the center of
the channel. As explained previously, for other values of $\eta$ the shape of
the finger is the same as that presented above, only its velocity changes.
\bifi[h]
	\centering
		\includegraphics[width=0.35\textwidth]{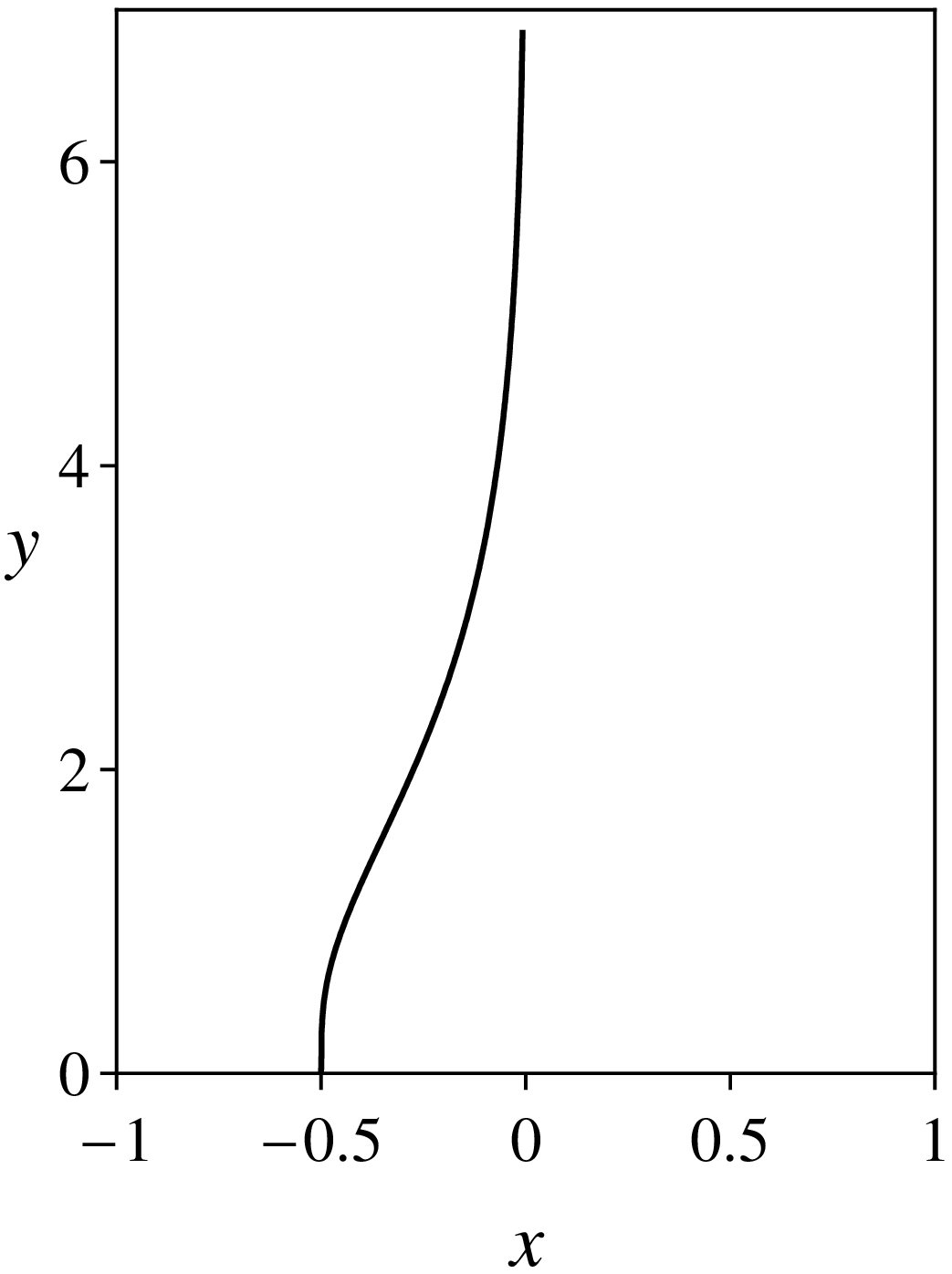}
	\caption{The growth of the single finger in the channel with $a(0)=-0.5$}
	\label{onefinger}
\efi
Naturally, if the initial position of the finger is already in the middle of the channel,
$a(0)=0$, it would simply continue growing along the centerline. In that case
the map $f_t$ is given by
\begin{equation}
f_t (z) =  \frac{2}{\pi} \arcsin \left( \sqrt{      \sin^2 \left(\frac{ \pi}{2} z
\right)
\cosh^2 \left(\frac{\pi}{2} H(t)\right)-\sinh^2 \left(\frac{ \pi}{2} H(t) \right)    }
\right),
\end{equation}
where $H(t)$ is the height of the finger at a given moment of time. In particular,
\bee
	|f''_t (0)| =  \frac{\pi}{2}  \coth \left( \frac{\pi}{2} H(t) \right),
\label{eq:f_bis}
\eee
which gives the velocity of the tip as
\bee
v(H)=\left( \frac{\pi}{2}  \coth \left( \frac{\pi}{2} H \right) \right)^{-\eta/2}.
\label{veloh}
\eee
The asymptotic velocity of the finger is thus
\bee
v_{as}=\lim_{H \rightarrow \infty} v(H) =
\left( \frac{\pi}{2} \right)^{-\eta/2},
\label{vas}
\eee
and the asymptotic growth factor is (cf. Eq.~\eqref{eq:dlaplace})
\bee
d_{as}= \left( \frac{\pi}{2} \right)^{-\eta/2-1}.
\label{das}
\eee
It is instructive to compare this result with that for a single finger growing in the half
plane. In the latter case
\begin{equation}
f_t (z) = \sqrt{z^2-H^2(t)}, \nonumber
\end{equation}
and the velocity obeys
\bee
v(H) = H^{\eta/2},
\label{hlfplane}
\eee
hence, for $\eta>0$, it is growing indefinitely as the finger
increases its height. This is to be expected since in the unbounded case, as the finger is
getting further away from the absorbing wall, it intercepts an increasingly larger flux.
Contrastingly, in the case of the channel, the total flux through its {cross-section} is
finite, and the growth rate of the finger stabilizes as soon as its height becomes large
in comparison to the channel width. Note that the result~\eqref{hlfplane} for the
{half-plane} may
also be recovered by expanding the formula~\eqref{veloh} in $H \ll 1$, since
$\frac{\pi}{2}  \coth \left( \frac{\pi}{2} H \right) = H + \dots$\ . Thus, at the
beginning
of the evolution, the finger behaves as if it were in an unbounded domain;
soon,
however, the presence of the side walls becomes a determining factor in its dynamics.

\section{From the channel to the cylinder}\label{chacyl}

The above formalism may also be used to find the evolution of the fingers in a channel
with
periodic boundary conditions:
\bee
u(x+2,y)=u(x,y),
\eee
topologically equivalent to the surface of a {semi-infinite} cylinder. Namely, consider a
single finger growing in the periodic channel with the initial position of the pole
$a(0)=a_0$. Translating the origin of the coordinate system to $a_0$ we obtain a single
finger growing vertically in the middle of the channel on the side walls of which both
periodic and reflecting boundary conditions are satisfied simultaneously. The growth of
this finger is described by the Loewner equation~\eqref{eq:loewnerP} with $a=0$.
Transforming back to the original coordinates one obtains
\bee
\dot{g}_{t} = d \frac{\pi}{2} \cot\left(\frac{\pi}{2}
(g_{t}-a) \right),
 \label{lcyl}
\eee
which is the Loewner equation for a single finger growth in the cylinder. This equation
has already been derived \cite{Bauer:2006,CarlesonMakarov:2002} in the context of growth
processes
in radial geometry (which may be then mapped onto cylindrical by the map of the
form $z \rightarrow \log z$).
However, yet another, and perhaps easier, derivation of~\eqref{lcyl} may be given,
starting form the Loewner equation for the {half-plane}~\eqref{eq:loewner} and summing
over
the periodic images of the pole:
$a_j=a+2j, \ j \in \mathds{Z}$. This gives
\bee
\dot{g}_{t} = \sum_{j=-\infty}^{\infty} \frac{d}{g_t-(a+2j)} = d \frac{\pi}{2} \cot\left(
\frac
{\pi}{2}
(g_{t}-a) \right).
\eee

The generalization of the Loewner equation to the {$n-$finger} case proceeds along the
same
lines as before and yields
\bee
\dot{g}_{t} = \sum_i d_i \frac{\pi}{2} \cot\left(\frac{\pi}{2}
(g_{t}-a_i) \right),
\label{ncyl}
\eee
whereas the equation of motion of the poles reads
\bee
\dot{a}_j = \sum_i d_i \frac{\pi}{2} \cot\left(\frac{\pi}{2}
(a_j-a_i) \right). \label{lcylpoles}
\eee
This time the self term~\eqref{eq:aprimP} is absent since the situation is again symmetric
and the {single-finger} slit mapping does not affect the position of the corresponding
pole.

\section{From the cylinder to the channel}\label{cylcha}

In the previous section we derived the dynamics of the fingers in the cylinder based
on the Loewner equation for the channel. Here we will go in the opposite direction and
derive the Loewner equation for the channel starting with Eq.~\eqref{lcyl}. As it turns
out, such an approach provides us with the clear interpretation of the self term
\eqref{eq:aprimP}.

\bifi
	\centering
	\includegraphics[width=0.5\textwidth]{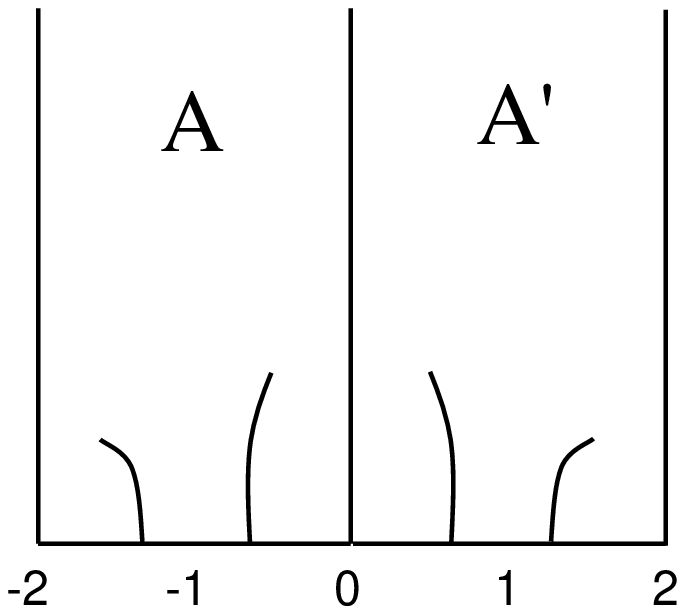}
\vspace{-0.5cm}
	\caption{The channel $A$ and its mirror reflection $A'$.}
	\label{dwapaski}
\efi

To start with, consider $n$ fingers growing in a channel with reflecting boundaries
(designated by $A$ in Fig.~\ref{dwapaski}). Next, let us reflect the system with respect
to
one of its side walls and denote the image by $A'$. Thus, there are now $2n$ fingers: the
original ones $\{1,\dots,n\}$ in $A$ and their images $\{n+1,\dots,2n\}$ in $A'$.
Due to the reflection symmetry
\bee
a_i=-a_{n+i}, \ \ \ \ \ d_i=d_{n+i}, \ \ \ \ \ \ \ i \in \{1 \dots n\},
\label{symmetry}
\eee
with coordinates chosen so that $x=0$ corresponds to the joint wall of the two channels,
as in Fig.~\ref{dwapaski}. Finally, we look for the solution  of the Laplace equation in
the joint system (AA') in
periodic boundary conditions. Due to the symmetry of the problem, such a solution will
automatically satisfy the reflecting boundary conditions in $A$.

To be more formal, the Loewner equation in the $AA'$ channel will be given by
\begin{equation}
\dot{g}_{t} = \frac{\pi}{4} \sum_{i=1}^{2n}  d_i \cot\left(\frac{\pi}{4}
(g_{t}-a_i) \right),
\label{1st}
\end{equation}
which is equivalent to~\eqref{ncyl} additionally rescaled by a factor of 2 to account for
the fact that
the width of $AA'$ is twice the width of $A$. Taking into account~\eqref{symmetry}, we
get
\begin{equation}
\dot{g}_{t} = \frac{\pi}{4} \sum_{i=1}^{n}  d_i \left[ \cot\left(\frac{\pi}{4}
(g_{t}-a_i) \right) + \cot\left(\frac{\pi}{4}
(g_{t}+a_i) \right) \right],
\end{equation}
which, by straightforward trigonometry, leads to
\begin{equation}
\dot{g}_{t} = - \sum_{i=1}^n  d_i \frac{\pi}{2} \frac{\sin\left(\frac
{\pi}{2}
g_{t} \right)}{\cos\left(\frac{\pi}{2} g_{t} \right) - \cos\left(\frac{\pi}{2}a_i
\right)}.
\label{2nd}
\end{equation}
Finally, we move the origin of coordinates to the center of $A$ by
\bee
z'=z+1,
\label{coord}
\eee
 which leads us to the Loewner equation for the channel~\eqref{Loewnern}.

The equation of motion of the poles may be transformed along the similar lines:
\begin{equation}
\dot{a}_j =   \frac{\pi}{4} \cot\left(\frac{\pi}{4}(
a_j-a_{n+j}) \right)  + \mathop{\sum_{i=1}^{2n}}_{i \neq j,n+j}  \frac{\pi}{4} \cot\left
(\frac{\pi}{4}(
a_j-a_i) \right).
\end{equation}
The first term describes the interaction between the pole $a_j$ and its image $a_{n+j}$
and, using~\eqref{symmetry} and~\eqref{coord}, may be transformed to $
- \frac{\pi}{4} d_j \tan \left(\frac{\pi}{2}a_j \right) $,
which is exactly the self term \eqref{eq:aprimP} derived before. The terms in the
remaining sum, using~\eqref{symmetry}, may be written in the form
$\cos (\frac{\pi}{2} a_{j}) {\displaystyle / } (\sin (\frac{\pi}{2} a_{j})   -
\sin (\frac{\pi}{2}a_{i}))$, in agreement
with~\eqref{eq:aprimP_n}.

The above derivation shows, in particular, that the stable, symmetric solution of {two-
finger} growth
in
the cylinder may be reduced to the previously considered problem of a single finger growth
in the channel. Indeed, Fig.~\ref{cylinder} shows {two-finger} solution in the cylinder
the
initial conditions
$a_{1,2}(0)=\pm 0.25$ (if the fingers are initially not symmetric with respect to zero
they can be rendered so by an
appropriate change of variables). It is observed that the {right-hand} side finger has the
same
shape
as that in Fig.~\ref{onefinger}, whereas the {left-hand} side one is its mirror
reflection.

\bifi[ht]
	\centering
		\includegraphics[width=0.35\textwidth]{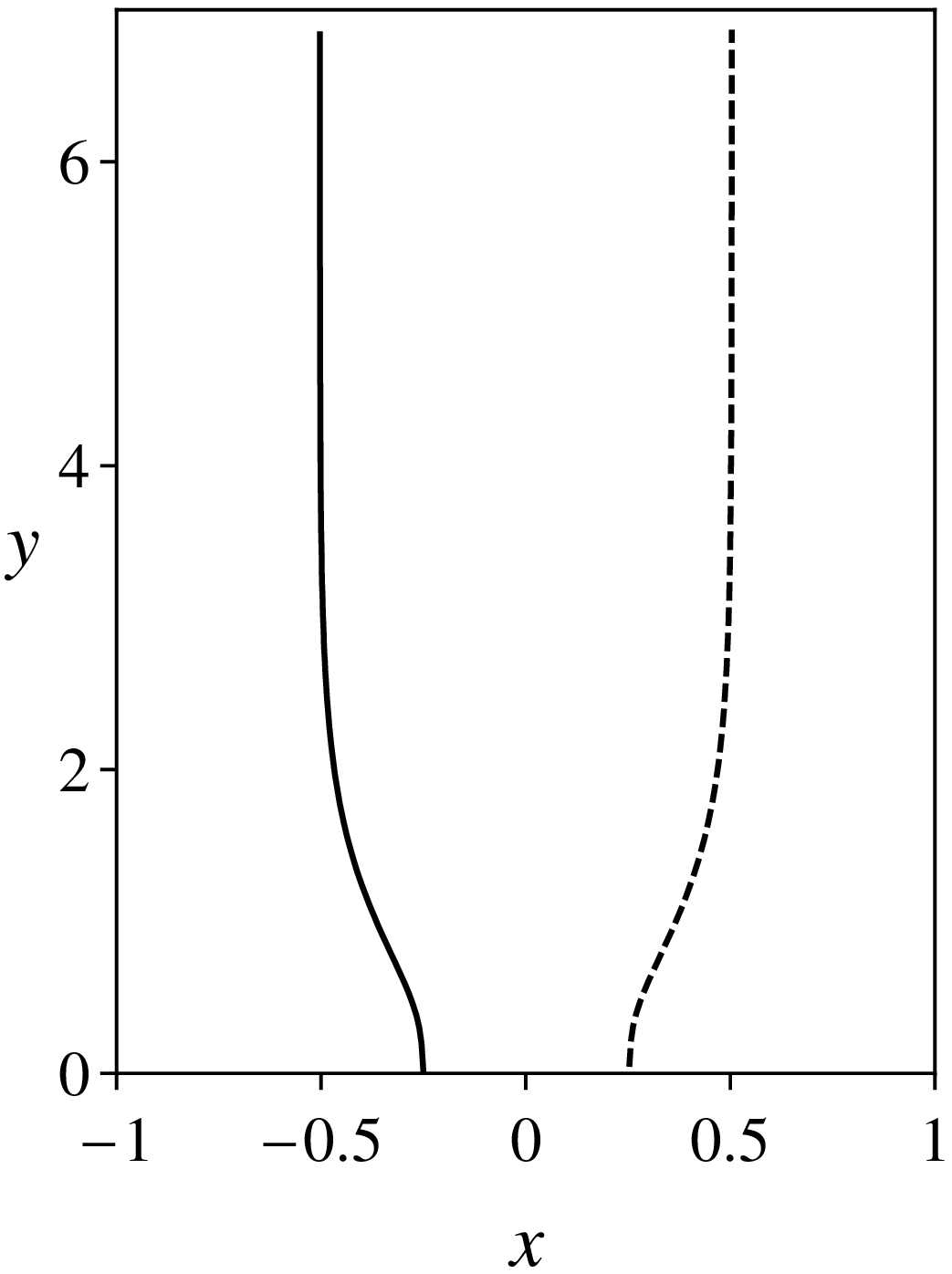}
		\caption{Two fingers growing in the cylinder with $a_{1,2}(0)=\pm 0.25$
and
$\eta=-2$ (stable symmetric solution).}
		\label{cylinder}
\efi

Fig.~\ref{cylinder} illustrates an important property of stable solutions of $n$ fingers
growing in the cylinder. Namely, irrespectively of initial positions, the fingers end up
in the symmetric configuration, with equal distances between each other. This is a direct
consequence of the repulsion between the poles, as described by Eq.~\eqref{lcylpoles}.
Indeed, it is seen that the uniform arrangement of the poles constitutes a fixed point
{of~\eqref{lcylpoles}}.

Because of the relation between the geometry of a reflecting channel and
that of a cylinder, the above statement may also serve to prove that, in the stable case,
both the fingers and the poles in the channel end up at the positions $x_i=(2i-1)/n-1$
where $n$ is the
number of fingers and $i=1,\dots,n$.

\section{The competition between fingers}

In this section we look more closely at the instabilities in the growth of two fingers
both in the channel and in the cylinder, paying particular attention to the screening
process and competition between the fingers. The results presented in this section,
unlike those presented in Figs.~\ref{onefinger} and~\ref{cylinder}, are obtained by
numerical calculation (as described in Appendix~\ref{numer}).

\bifi[ht]
	\centering
		\includegraphics[width=0.55\textwidth]{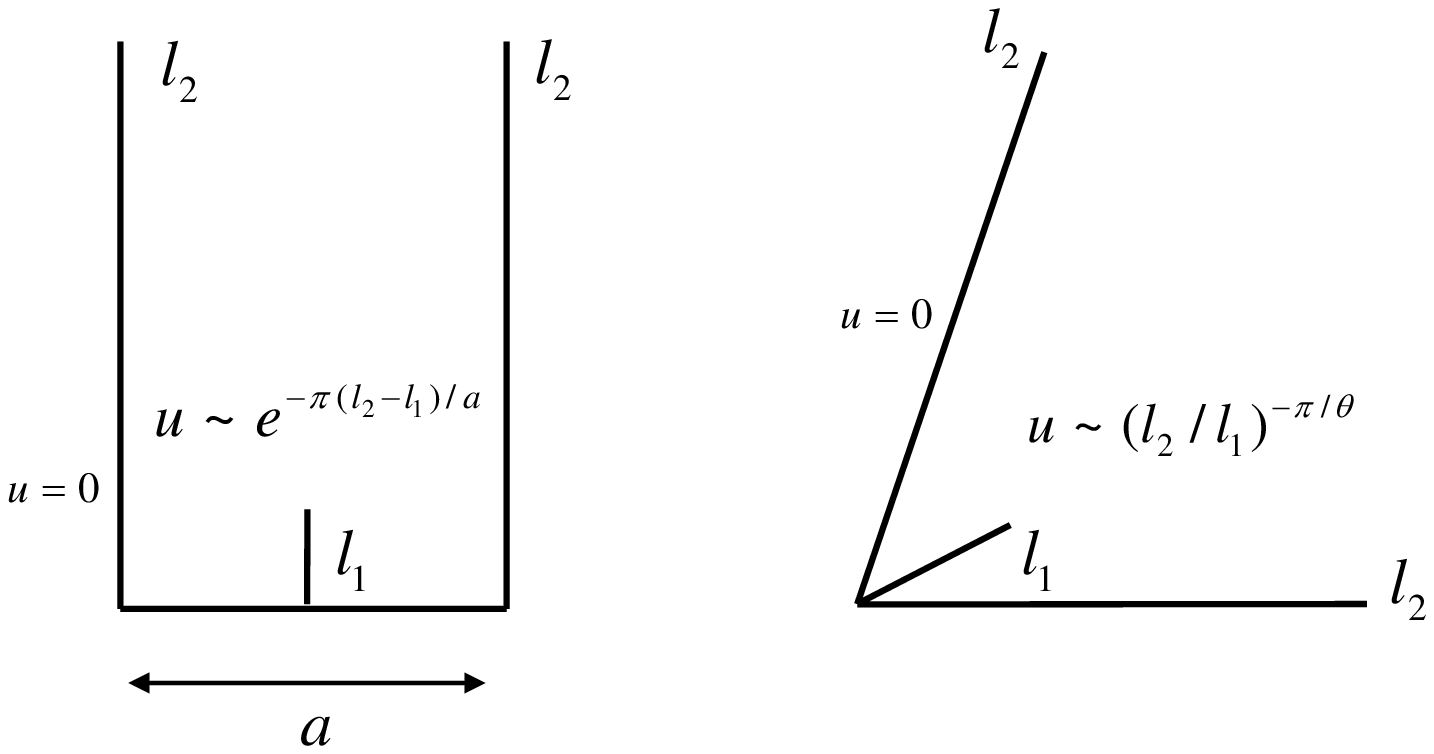}
                \vspace{-1cm}
		\caption{The scaling of a Laplacian field $u$ in rectangular and
{wedge-like} fjords.}
		\label{fjords}
\efi

As mentioned in Sec.~\ref{hplane}, in the case where fingers grow in the half-plane, there
exists a critical value of the exponent ($\eta_c$) below which the symmetric solution is
stable nad another threshold value, $\eta_c^{\prime}$, such that for
$\eta_c < \eta < \eta_c^{\prime}$, there exists an asymmetric solution of the
growth of two fingers with the velocity ratio of the slower finger to that of the faster
one, $\frac{v_1}{v_2}$, asymptotically
approaching the value different both from 0 and 1.  Here we argue that for the channel,
either reflecting or periodic,
$\eta_c=\eta_c^{\prime}=0$, i.e. the growth is unstable for any positive $\eta$, with
$\frac{v_1}{v_2} \rightarrow 0$. Such a
fundamental difference between
the two systems is connected
with the geometries involved~\cite{Evertsz:1991}. Consider the late stages of the
screening process, with a shorter finger (of length $l_1$) situated in a deep fjord
between two
longer fingers (or a longer finger and its periodic image) of length $l_2$. The value of
the Laplacian field $u$ near the tip of the short finger then obeys
\bee
u \sim e^{- \pi (l_2-l_1)/a}, \ \ \ \ \ \ \  l_2 \gg l_1,
\label{fjord1}
\eee
where $a$ is the width of the fjord.
As mentioned, in the {half-plane} case the fingers asymptotically tend to straight lines
growing  radially from the origin, thus in that case the fjords will have a {wedge-like}
shape (cf. Fig.~\ref{fjords}). For such a fjord with an internal angle $\theta$ one
finds
\bee
u \sim \left(\frac{l_2}{l_1} \right)^{-\pi/\theta}, \ \ \ \ \ \ \  l_2 \gg l_1.
\label{fjord2}
\eee
Thus the screening in the channel geometry is exponential, whereas in the {half-plane} it
follows a {power-law}. Consequently, in the latter case, we can choose the values of
$\eta$ and $\theta$  in such a way that would guarantee that the ratio of the velocities
of the fingers is equal to their length ratio
\bee
\frac{l_1}{l_2} = \frac{v_1}{v_2},
\eee
which corresponds to the (asymmetric) solution of the fingers dynamics which is stationary
in the sense that the ratio $l_1/l_2$ remains constant.

Contrastingly, for the exponential screening, irrespectively of the values of $a$ and
$\eta$ (as long as it is positive) the velocity of the shorter finger always decreases to
zero, thus also $l_1/l_2 \rightarrow 0$: the screening is complete.

\bifi[ht]
	\centering
		\includegraphics[width=0.6\textwidth]{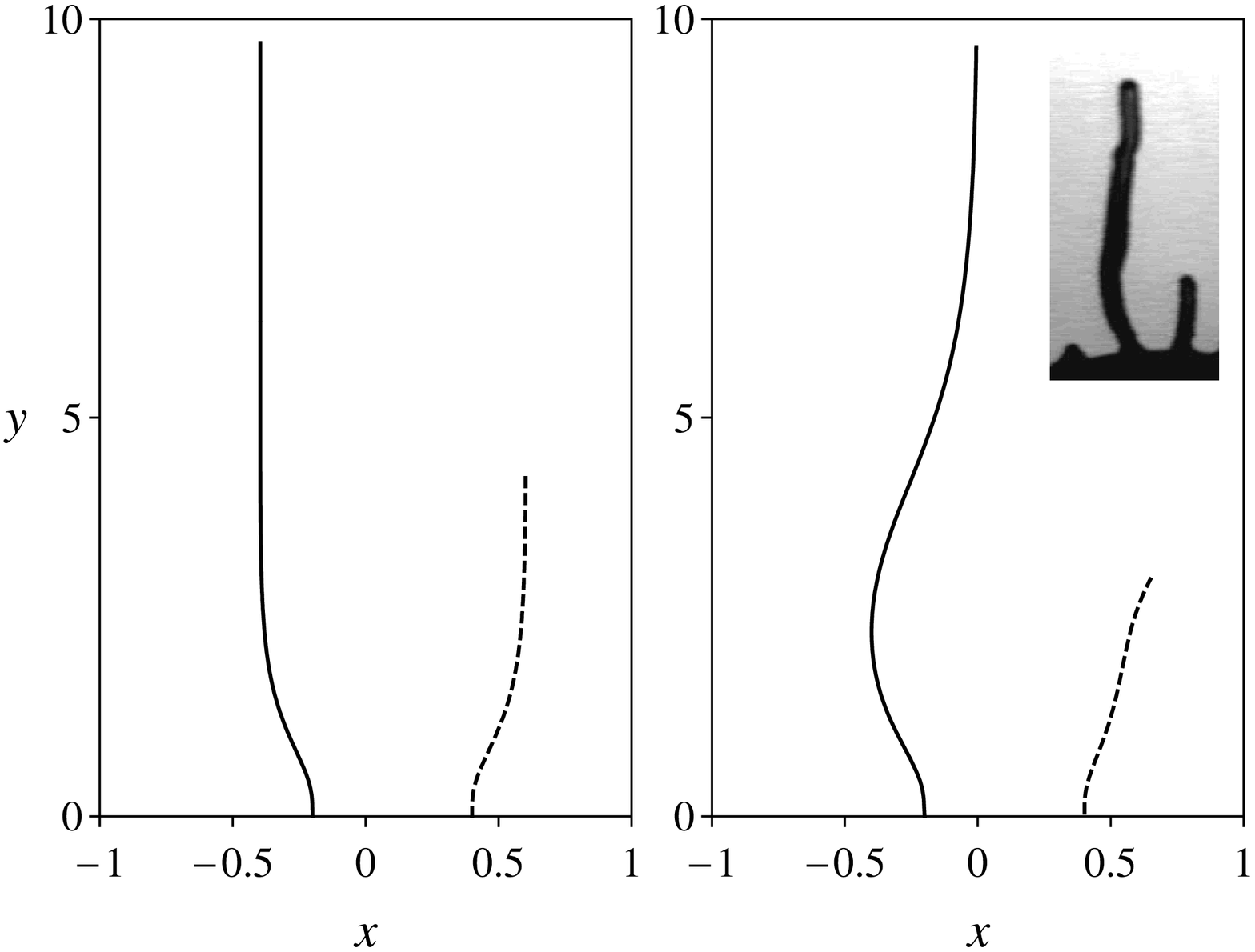}
		\caption{Two fingers growing in the cylinder (left panel) and in the
reflecting channel (right panel) for $\eta=1$. The initial
positions of the fingers are $a_1(0)=-0.2$ and $a_2(0)=0.4$. Inset: the interaction of two
fingers in the combustion experiments of Zik and Moses~\cite{Zik:1999}.}
		\label{fingers}
\efi

Fig.~\ref{fingers} presents an example of such a situation. Here, two fingers are evolving
in the channel with either reflecting or periodic walls at $\eta=1$. At the beginning, the
fingers repel each other and behave very similarly to the stable, symmetric solution with
$\eta<0$ (cf. Fig.~\ref{together}). However, as soon as the their height becomes
comparable to the width of the channel, the instability sets in and initially small
differences in height of the fingers are rapidly amplified. As observed in
Fig.~\ref{fingers}, in the reflecting channel the longer (winning) finger is attracted to
the centerline (as it is the case for the single finger solution), whereas in the cylinder
it
continues to grow vertically.

\bifi[ht]
	\centering
		\includegraphics[width=0.37\textwidth]{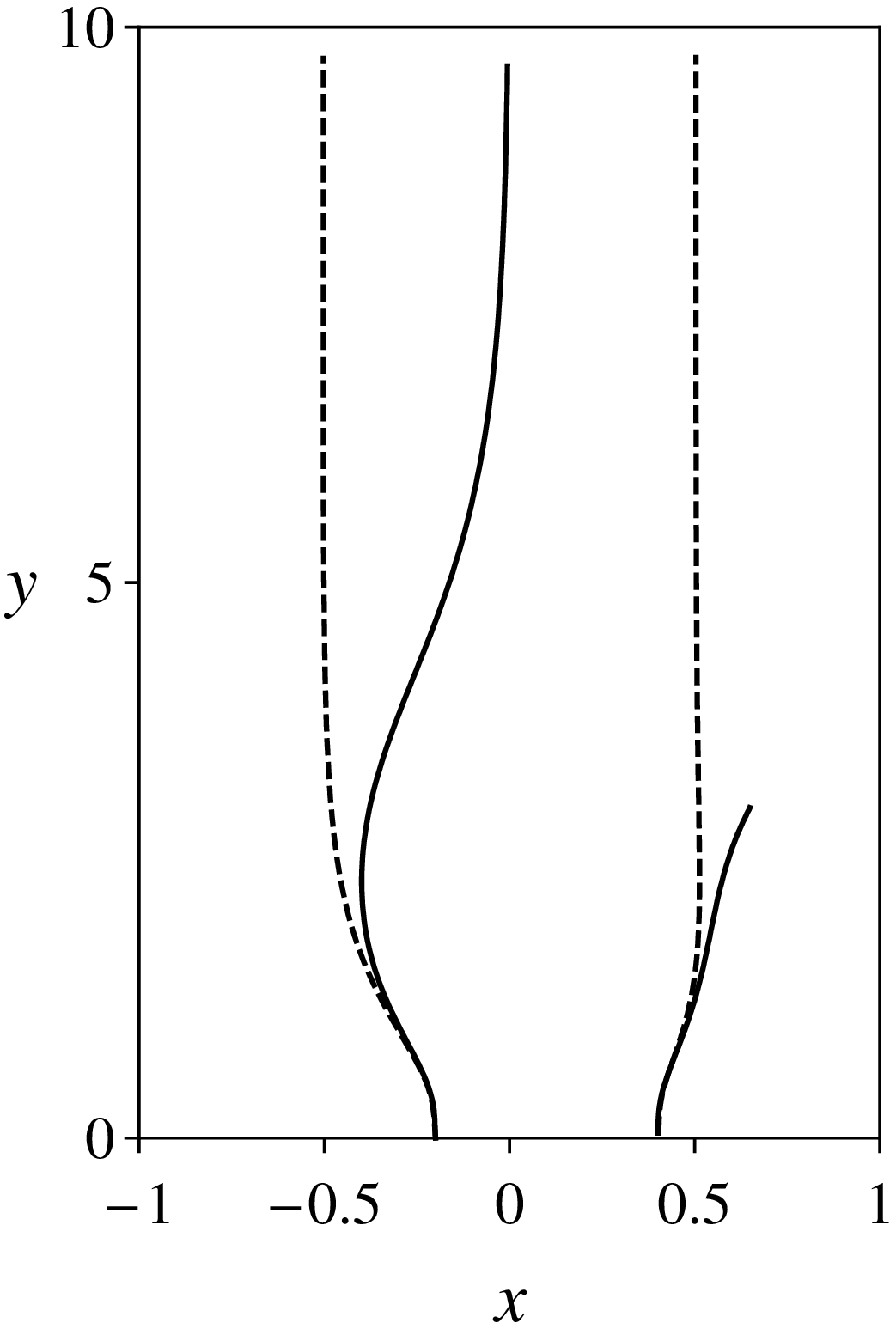}
		\caption{Two fingers growing the reflecting channel at $\eta=1$ (solid
line) and $\eta=-2$ (dashed). The initial
positions of the fingers are the same as those in Fig~\ref{fingers}.}
		\label{together}
\efi

The analysis of Fig.~\ref{fingers} together with the corresponding dependencies of the
growth
velocity on time presented in Fig.~\ref{velocities} shows distinctly three different
stages
of
finger growth. In the
initial stage, the fingers repel each other, their velocities increase quickly but remain
equal to each other; the competition between the fingers is not yet present.
Then, as the distance between the fingers approaches 1, their
velocities stabilize at around $v \approx 0.6$. In fact, the velocities of the fingers in
this stage
are close to the asymptotic velocity of a single finger in the channel of half the width
of
the original one, i.e., analogously to Eq.~\eqref{vas}
\bee
v = \left(\frac{\pi}{4}\right)^{-1/2} \approx 0.56,
\eee
which agrees with the value $0.6$ quoted above. As seen in Fig.~\ref{together} the
unstable $\eta=1$ solution diverges from the stable one ($\eta=-2$) when the height of the
finger reaches approximately $1-2$ units, i.e. becomes comparable
with the channel width. This is the moment where the final, third stage of the evolution
begins, accompanied by the sharp decrease increase of the velocity of the losing finger
and the corresponding increase of the velocity of the winning one up to the asymptotic
value given, according to \eqref{vas}, by $(\pi/2)^{-1/2} \approx 0.798$.

Note that in the case of the cylinder it is impossible to predict
beforehand which finger is expected to win the competition -- the initial situation is
fully symmetric and in principle the system could remain in the unstable symmetric state
were it not for the numerical noise (no other sources of noise are present).
Contrastingly,
in the reflecting channel, the finger which starts closer to the centerline is destined
the
win, since, during its evolution it moves through regions of higher field gradient. This
is
also the reason why the instability sets in earlier in the case of the
channel.

\bifi[ht]
	\centering
		\includegraphics[width=0.9\textwidth]{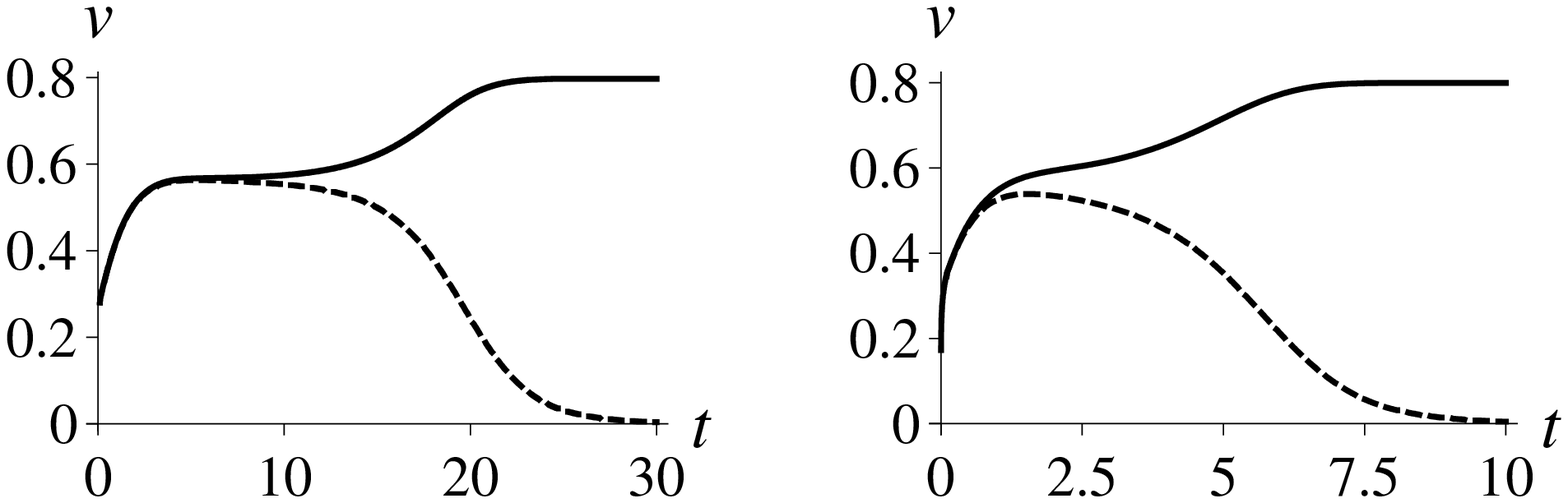}
		\caption{The velocities of the fingers as a function of time in the
cylinder (left) and in the channel (right),
corresponding to the fingers of Fig.~\ref{fingers}.}
		\label{velocities}
\efi

\bifi[ht]
	\centering
		\includegraphics[width=0.9\textwidth]{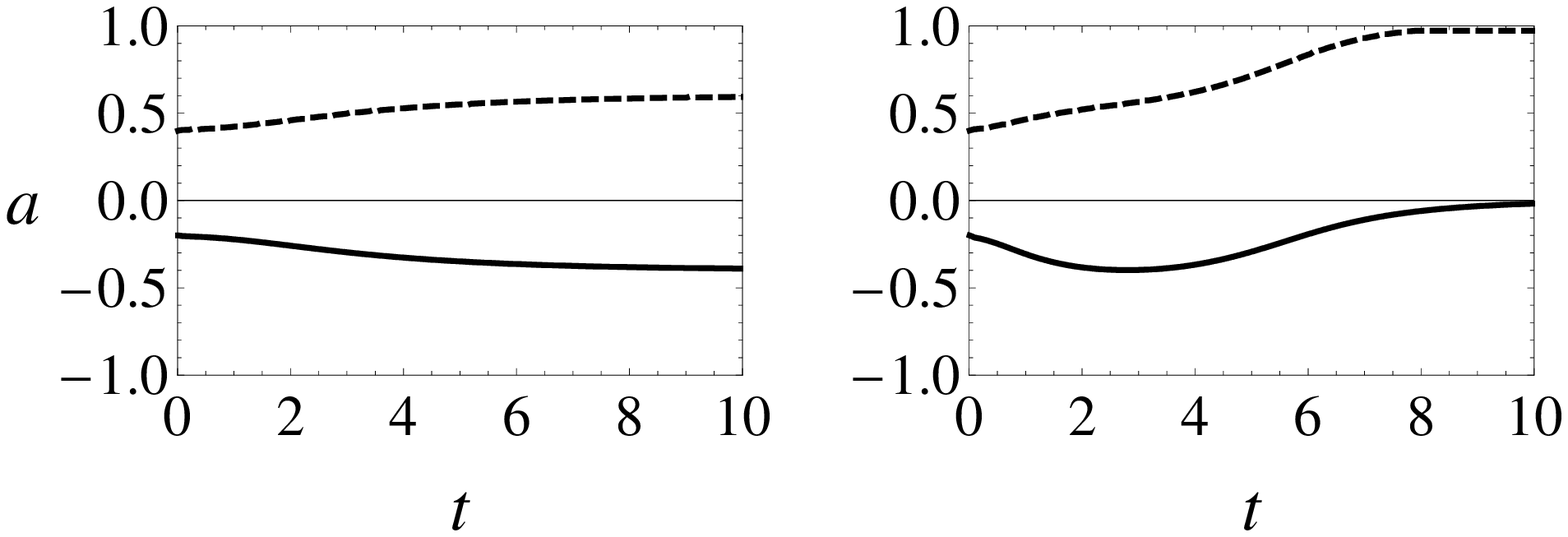}
		\caption{The positions of the poles as a function of time, in the cylinder
(left) and in the channel (right), corresponding
to the fingers of Fig.~\ref{fingers}.}
		\label{poles}
\efi

Further insight into the competition process may be gained by analyzing the evolution of
the
poles, as presented in Fig.~\ref{poles}. It is seen that in the channel the pole of the
winning finger is
attracted towards the center, whereas the pole of the suppressed one moves to the side of
the channel. Such a behavior of the poles may be explained based on the evolution
equation~\eqref{eq:aprimP_n} in which the growth factors $d_i$ have been replaced by their
asymptotic values: $d_1=d_{as}$ (cf. Eq.~\eqref{das}) for the winning finger and
$d_2=0$ for the losing one. This leads to
\bee
	\dot{a}_1 &=& - \frac{\pi}{4} d_{as} \ \tan\left(\frac{\pi}{2}a_1 \right),
\nonumber
\\
	\dot{a}_2 &=& \frac{\pi}{2} d_{as}  \frac{\cos\left(\frac{\pi}{2} a_2 \right)}{
\sin
\left(\frac{\pi}{2} a_2 \right) - \sin\left(\frac{\pi}{2} a_1 \right)} .
\eee
The asymptotic fixed point of the first equation is $a_1=0$, in full analogy to the single
finger solution in the channel. It is then inserted into the second equation to yield
\bee
	\dot{a}_2 = - \frac{\pi}{4} d_1 \ \cot \left(\frac{\pi}{2} a_2 \right)
\eee
with the asymptotic fixed points at $a_2=\pm 1$.

In the inset of Fig.~\ref{fingers}, we reproduce a {two-finger} pattern from the
combustion
experiments by Zik and Moses. The resemblance is remarkable. Nevertheless, we do not claim
that such a simple model can accurately reproduce the complex dynamics of combustion in
{Hele-Shaw} cell, we can only conclude that the fingers in those experiments indeed seem
to
follow the gradient lines of the Laplacian field, (which in this case represents the
concentration of oxygen at a given point) and that there is a screening mechanism leading
to
the competition between the fingers. This does not mean necessarily that the velocity of
the
finger must be connected with the field gradient analogously to Eq.~\eqref{veloc}. In
fact,
many of the results of the above presented model are independent of the actual relation
between the velocity $v$ and the field $u$.
In particular, the shape of the {single-finger} solution and the symmetric {two-finger}
solution are independent of $v(u)$ relation, which affects in this case only the overall
growth rate, not the finger shape. The same does not hold for asymmetric solutions, but
even
in that case some of the qualitative features of the dynamics, which give the winning
finger in
{Fig.~\ref{fingers}} its characteristic shape remain universal: at the beginning the
fingers
are
attracted to the symmetric solutions, then the instability sets in and finally the winning
finger is attracted to the centerline.

\section{Summary}
In this paper we have studied a simple model of Laplacian growth,
in which the {finger-like} protrusions grow only at their tips. In contrast to the
{needle-growth} models, considered
previously in the literature, \cite{
Peterson:1989,PetersonFerry:1989,Kurtze:1991,DerridaHakim:1992,Peterson:1998,Krug:1993,
Huang:1997,Bernard:2001,Sakaguchi:2007}, here the fingers are allowed to bend along the
gradient of the Laplacian field. The method of iterated conformal maps  was applied to
solve
the growth problem in the geometry of the channel with two reflecting walls. We derived
the
Loewner equation for such a geometry and analyzed its analytical and numerical solutions
in
few simple cases. It was shown that the finger growth in the channel is qualitatively
different from that in the unbounded space. In particular, the competition between the
fingers is much stronger in the case of the channel and, for any positive value of the
exponent $\eta$, the only stable asymptotic situation is that of a single finger, which
had
outgrown the others and continues to grow  with a constant velocity along the centerline
of
the channel, whereas the other fingers stop growing completely.

\acknowledgments

Valuable discussions with Adam Szereszewski are gratefully acknowledged. The photos of the
combustion experiments  are courtesy of prof. Elisha Moses from the Weizmann Institute of
Science. This project has been supported by the Polish Ministry of Science and Higher
Education (Grant No. N202023 32/0702).

\appendix

\section{Expression for finger velocity in terms of the map ${\bf f_t}$}\label{efbis}

In this appendix a sketch of the proof of Eq.~\eqref{eq:v_f} is presented. This equation
relates the finger velocity to $f_t''(a)$ -- the second derivative of $f_t$ mapping
calculated
at the image of the tip. To prove ~\eqref{eq:v_f}, it is convenient to introduce the
analytic complex potential $\Theta$, such
that $u=\text{Im} \Theta$. Then, in the $\omega$ plane, corresponding to the empty
{half-plane} (or channel), the
solution satisfying the boundary condition $u(\omega)=0$ on the real axis is
$\Theta(\omega) = \omega$. The corresponding complex potential in the $z$ plane is then
simply $\Theta(g_t(z))= g_t(z)$. The derivative of the complex potential is directly
connected
to
the gradient of $u$, in particular
\bee
\left| \frac{dg_t}{dz}\right| = \left| \nabla u \right|.
\eee
Ignoring for the moment the singularity at the tip
{($z=\gamma$)},
let us try to calculate the derivative of the $g$ at that point
\bee
\left. \frac{dg_t}{dz}\right|_{z=\gamma} \hspace{-0.2cm} = \lim_{\delta \rightarrow 0}
\frac{g_t(\gamma+\delta)-g_t(\gamma)}{\delta}=
\lim_{\delta \rightarrow 0} \frac{a+\epsilon(\delta)-a}{\delta} = \lim_{\delta \rightarrow
0} \frac{\epsilon(\delta)}{\delta},
\eee
where $a+\epsilon$ is the image of the point $\gamma+\delta$ under $g_t$.

On the other hand
\bee
\delta = f_t(a+\epsilon) - f_t(a) = f_t'(a) \epsilon + \frac{1}{2} f_t''(a) \epsilon^2 +
\dots \quad .
\eee
The first term on the right hand side vanishes, since $f$ has a local maximum at $a$,
corresponding to the tip of the finger, thus $\epsilon = (2 \delta / f_t''(a))^{1/2}$ and
the expression for the gradient takes form
\bee
\left. \frac{dg_t}{dz}\right|_{z=\gamma}  \hspace{-0.2cm}  = \lim_{\delta \rightarrow 0}
(\delta f_t''(a) /2)^{-1/2},
\eee
\noindent with the singularity at $\delta=0$, as expected. Using the regularization
described in
Sec.~\ref{themodel} to remove the factor $\delta^{-1/2}$ we get finally
\bee
|v| \sim \left| f_t''(a) \right|^{-\eta/2}.
\eee

\section{Numerical method}\label{numer}

In this appendix, we describe briefly the numerical method used to follow the evolution of
the fingers.
Instead of integrating the Loewner equation, as it is done e.g. in
Ref.~\cite{Kager:2004}, we obtain the mapping $g_t$ by direct iteration of the elementary
slit mappings (which are of the form~\eqref{phiz} for the half-plane and ~\eqref{eq:phiP}
for the channel geometry). Since there are $n$ fingers, each timestep
involves the composition of $n$ slit mappings, $\phi_i$, each characterized by a
corresponding position of the pole, $a_i$, and the growth factor $d_i$. In the case of the
cylinder, the corresponding slit mapping may also be obtained from Eq~\eqref{eq:phiP} by a
procedure described in Sec.~\ref{chacyl}: for a finger with the pole at $a_i$ first the
system of coordinates is translated by $z \rightarrow z-a_i$, then the slit mapping
$\eqref{eq:phiP}$ is applied (the corresponding pole is now located at the origin, thus
$a=0$ is to be put in Eq.~\eqref{varphi}) and the result is transformed back to the
original coordinate
system:
$z \rightarrow z+a_i$.

The calculations are somewhat complicated by the dependence of the growth factors $d_i$ on
the $f_t$ mapping, as given by Eq.~\eqref{eq:dlaplace}. However, as before, this map may
also
be obtained by the composition of elementary mappings $\tilde{\phi}$, which are the
inverses
of slit mappings $\phi$, i.e. $\tilde{\phi}(\phi(z))=z$. The second derivative of $f$ may
then be calculated either analytically, by differentiating the composition of all the slit
mappings which make up $f_t$, or by a direct numerical differentiation. The latter method
is
faster, whereas the former is more accurate which becomes important at the late stages of
finger competition, where the growth factor of the losing finger becomes very small.

\bibliography{loewner}

\end{document}